\documentclass[useAMS,usenatbib,onecolumn,usegraphicx]{mn2e}

\newcommand{\gsim}{\, \raisebox{-0.8ex}{$\stackrel{\textstyle >}{\sim}$ }}
\newcommand{\lsim}{\, \, \raisebox{-0.8ex}{$\stackrel{\textstyle <}{\sim}$ }}

\newcommand{\beq}{\begin{equation}}
\newcommand{\eeq}{\end{equation}}
\newcommand{\beqar}{\begin{eqnarray}}
\newcommand{\eeqar}{\end{eqnarray}}

\usepackage{ulem}
\usepackage{color}

\newcommand{\Ibar}{I\hspace{-.5em}{\scriptscriptstyle \stackrel{\mathbf{-}}{}}}

\title[Avoided crossing in GWs from PNS]  % up to 40
{Avoided crossing in gravitational wave spectra from protoneutron star} 
\author[H. Sotani \& T. Takiwaki]
{Hajime Sotani$^{1,2}$ \thanks{E-mail:sotani@yukawa.kyoto-u.ac.jp} and
Tomoya Takiwaki$^{3,4}$
\\
$^1$Astrophysical Big Bang Laboratory, RIKEN, Saitama 351-0198, Japan\\
$^2$Interdisciplinary Theoretical \& Mathematical Science Program (iTHEMS), RIKEN, Saitama 351-0198, Japan\\
$^3$Division of Science, National Astronomical Observatory of Japan, 2-21-1 Osawa, Mitaka, Tokyo 181-8588, Japan\\
$^4$Center for Computational Astrophysics, National Astronomical Observatory of Japan, 2-21-1 Osawa, Mitaka, Tokyo 181-8588, Japan}

\begin{document}
\maketitle
\label{firstpage}

%%%%%%%%%%%%%%%%%%%%%%%%%%%%%%%%%%%%%%%%%%%%%%%%
% Abstract (Main Journal 250 words max; Letters 200 words max.)
\begin{abstract}
The ramp up signals of gravitational waves appearing in the numerical simulations could be important signals to estimate parameters of the protoneutron star (PNS) at supernova explosions. To identify the signals with PNS oscillations, we make a linear perturbation analysis and compare the resultant eigenfrequencies with the ramp up signals obtained via the two-dimensional numerical simulations. Then, we find that the ramp up signals correspond well to the $g_1$-mode in the early phase and to the $f$-mode, to which the $g_1$-mode is exchanged via the avoided crossing. We also confirm that the $f$- and $g_1$-modes are almost independent of the selection of the PNS surface density in the later phase after core bounce. In addition, we successfully find that the fitting formula of $g_1$- and $f$-modes, which correspond to the ramp up signals in the numerical simulation, as a function of the PNS average density. That is, via the direct observation of the gravitational waves after supernova explosion, one could extract the time evolution of the PNS average density by using our fitting formula. 
\end{abstract}
%%%%%%%%%%%%%%%%%%%%%%%%%%%%%%%%%%%%%%%%%%%%%%%%

\begin{keywords}
stars: neutron  -- equation of state -- stars: oscillations
\end{keywords}

%%%%%%%%%%%%%%%%%%%%%%%%%%%%%%%%%%%%%%%%%%%%%%%%
\section{Introduction}
\label{sec:I}
%%%%%%%%%%%%%%%%%%%%%%%%%%%%%%%%%%%%%%%%%%%%%%%%

Owing to the success of the direct detection of gravitational waves from compact binary mergers, the gravitational waves are now a new tool to obtain an astronomical information. In particular, at the event GW170817, not only the gravitational waves but also the electromagnetic counterparts have been detected \citep{GW6,EM}, which signifies an advent of multi-messenger era, i.e., the gravitational waves become crucial for astronomy and astrophysics as well as the electromagnetic and neutrino signals. During the third observing run by the advanced LIGO (Large Interferometer Gravitational-wave Observatory) and advanced Virgo, additional gravitational wave events (and their candidates) have already been reported, where the Japanese gravitational wave detector, KAGRA \citep{aso13}, will also join in the network soon. Furthermore, the discussion for the third-generation gravitational wave detectors, such as the Einstein Telescope and Cosmic Explorer \citep{punturo,CE}, has already been started. Thanks to these attempts, one would expect to detect gravitational waves  in the near future from not only the compact binary mergers but also core-collapse supernovae, which correspond to the last moment of a massive star's life.

As in the case of the studies of compact binary mergers, the core-collapse supernovae have been studied basically by numerical simulations. Through these studies, it has been sometimes discussed how gravitational waves would be radiated from such a system (e.g., \cite{Murphy09,MJM2013,Ott13,CDAF2013,Yakunin15,KKT2016,Andresen16,Richers2017,Takiwaki2017,OC2018,RMBVN19,VBR2019,PM20}). The frequency of the gravitational wave signals reported in some of these studies increases with time after core-bounce from a few hundred hertz up to around kilohertz. This ramp up signal is initially considered as a result of the Brunt-V\"{a}is\"{a}l\"{a} frequency at the protoneutron star (PNS) surface \citep{MJM2013,CDAF2013}, which is often called as the surface gravity ($g$-) mode. Meanwhile, since the Brunt-V\"{a}is\"{a}l\"{a} frequency locally determined with the PNS properties is not generally associated with a global oscillations of the system, the ramp up signal appearing in the numerical simulations may come from a kind of the global oscillations of the PNS or core region of progenitor. Anyway, it is important to find an association between the gravitational wave signals and PNS properties if possible, where the gravitational waves would tell us an ``invisible" information of PNSs.

In order to extract a physical property of astronomical objects, asteroseismology is another powerful technique. This is a similar method in seismology for the Earth and helioseismology for the Sun, where one could see the property as an inverse problem by using the specific signals radiated from the objects. In particular, for the case of compact objects, gravitational waves are important for adopting this method, which is sometimes called as gravitational wave asteroseismology \citep{KS1999}. Up to now, the studies with asteroseismology for cold neutron stars have been done extensively. For example, the neutron star crust properties could be constrained by identifying the quasi-periodic oscillations observed in the giant flares with the crustal torsional oscillations (e.g., \cite{GNHL2011,SNIO2012,SNIO2013a,SNIO2013b,SIO2016,SIO2017,SIO2018,SIO2019}), or magnetic properties in neutron stars could also be understood by identifying with magneto(-elastic) oscillations (e.g., \cite{SKS2007,SKS2008,CK2012,GCFMS2011,GCFMS2013b}). It is also proposed that the radius, mass, and equation of sate (EOS) of compact objects would be  restricted via the direct detection of the gravitational waves (e.g., \cite{AK1996,AK1998,STM2001,SKH2004,SYMT2011,PA2012,DGKK2013}).

Unlike the situation for cold neutron stars, the studies with asteroseismology for PNSs are relatively poor and fresh, which may partially come from the difficulty for providing the PNS model as a background for the linear analysis. Actually, the cold neutron star models are easily produced by integrating the Tolman-Oppenheimer-Volkoff equation together with a relation between the pressure and density, while one has to know the radial distributions of pressure, density, electron fraction, and entropy per baryon for producing the PNS models. However, these radial distributions inside the PNSs can be determined in principle only as a result of numerical simulation for core-collapse supernovae. Even so, the number of studies about PNS asteroseismology is gradually increasing in past years by virtue of the development of the numerical simulations \citep{FMP2003,Burgio2011,FKAO2015,ST2016,Camelio17,SKTK2017,MRBV2018,SKTK2019,SS2019,ST2020,TCPF2018,TCPOF2019a,TCPOF2019b,WS2019}.

So far, there are mainly two approaches for examining the asteroseismology for PNSs, where the PNS model (or the numerical domain) for making a linear analysis is different, depending on the approach. One is the approach that the global oscillations inside the PNSs are examined, where only the PNS is considered and its surface is defined by a threshold value of $\rho_s$ \citep{FMP2003,Burgio2011,FKAO2015,ST2016,Camelio17,SKTK2017,MRBV2018,SKTK2019,SS2019,ST2020}. In this approach, the boundary condition at the outer boundary, i.e., the PNS surface, is the same as that in the standard asteroseismology, i.e., the Lagrangian perturbation of pressure should be zero at the PNS surface. We note that this boundary condition is just an approximation, because the PNS is surrounded by in-falling material during core-collapse supernovae. Then, as in the standard asteroseismology, the eigenmodes are identified by counting the nodal numbers in the eigenfunction. The disadvantage in this approach is the fact that the eigenfrequencies may depend on the selection of $\rho_s$, but at least some of eigenfrequencies, including the fundamental ($f$-) mode frequency, seem to be independent of the selection of $\rho_s$ at the late phase, such as $\sim 500$ ms after core bounce \citep{MRBV2018}. The second approach is that one considers the global oscillations inside the region up to the shock radius \citep{TCPF2018,TCPOF2019a,TCPOF2019b}. With this approach, by definition one can avoid the uncertainty about the position of outer boundary. On the other hand, the boundary condition at the outer boundary may not be well-discussed yet, which is assumed that the radial component of the Lagrangian displacement should be zero in  \cite{TCPF2018,TCPOF2019a,TCPOF2019b}. Since this boundary condition is completely different from that in the standard asteroseismology, one has to redefine the eigenmodes with a new criterion. In fact, the eigenmode is identified with considering the shape of the eigenfunction at the late phase after core bound in \cite{TCPF2018,TCPOF2019a,TCPOF2019b}. We remark that the discussion based on mode-function matching have been done recently \citep{WS2019}, in addition to the above two approaches. In this study, we simply adopt the first approach.

One of the most important tasks in PNS asteroseismology is the understanding of the ramp up signals of gravitational waves appearing in the numerical simulations. So, in this study we mainly compare the gravitational wave signals obtained in the two-dimensional (2D) numerical simulations to the eigenfrequencies excited in the corresponding PNS models, adopting the relativistic Cowling approximation. Then, we will discuss the origin of ramp up signals in numerical simulations, which is associated with the PNS properties. We should remark that a few studies have been done with taking into account some terms of metric perturbations  \citep{MRBV2018,TCPOF2019a}. That is, the perturbation of the lapse function is taken into account in \cite{MRBV2018}, where the frequencies (at least the $f$-mode) calculated with the Cowling approximation seem to be underestimated, while the perturbations of the lapse function and the conformal factor are taken into account in \cite{TCPOF2019a}, where the frequencies calculated with the Cowling approximation seem to be overestimated. Anyway, with the Cowling approximation, the calculated frequencies become real values, because one neglects an energy release due to the gravitational wave radiation, where one can not deal with the modes associated with the spacetime oscillation, i.e., the so-called $w$-modes.

This paper is organized as follows. In Sec. \ref{sec:PNSmodel}, we describe the PNS models considered in this study. In Sec. \ref{sec:Oscillation}, we show the eigenfrequencies of gravitational waves from the PNS, compare them to the spectra obtained from the numerical simulation, and discuss their dependence on the PNS properties. Then, we make a conclusion in Sec. \ref{sec:Conclusion}. Unless otherwise mentioned, we adopt geometric units in the following, $c=G=1$, where $c$ denotes the speed of light, and the metric signature is $(-,+,+,+)$.

%%%%%%%%%%%%%%%%%%%%%%%%%%%%%%%%%%%%%%%%%%%%%%%%
\section{PNS Models}
\label{sec:PNSmodel}
%%%%%%%%%%%%%%%%%%%%%%%%%%%%%%%%%%%%%%%%%%%%%%%%

In order to prepare the PNS models, first we did the one-dimensional (1D) and 2D hydrodynamical simulation. We remark that the 1D data is just for reference, which is considered only here and in Appendix \ref{sec:appendix_2}. The setup of neutrino radiation hydrodynamic simulations is basically the same as that of the previous work \citep{ST2020}, where {\small 3DnSNe} code is used and neutrino transport is solved by isotropic diffusion source approximation \citep{liebendoerfer2009,takiwaki2014}.
The code is also used in \cite{takiwaki2016,oconnor2018,kotake2018,nakamura2019,sasaki2019,zaizen2019,ST2020}, to solve 1D, 2D and three-dimensional (3D) hydrodynamic equations for core-collapse supernovae. We employ the resolution of the spherical polar grid of $512$ for 1D run and $512\times128$ for 2D run.
The radial grid ($r$) covers $0$ -- $5000\mathrm{~km}$ and polar grid ($\theta$) covers $0$ -- $\pi$.
The adopted set of neutrino reaction is the same as in \cite{kotake2018}.

In this study, we particularly adopt $2.9M_\odot$ He star (He2.9) without rotation as a progenitor model \citep{He29}, whose mass is relatively small because it is a model without hydrogen outer layer. With this progenitor mode, we adopt LS220 \citep{LS220} as the EOS in a high density region, which is constructed with the compressible liquid drop model with the incompressibility $K_0=220$ MeV and the slope parameter of symmetry energy $L=73.8$ MeV. We remark that the maximum mass of a cold neutron star constructed with LS220 is $2.0M_\odot$. The 2D model explodes at 280 ms postbounce when the averaged shock radius reaches 400km. On the other hand, the 1D model does not explode within 1000 ms postbounce.

\begin{figure*}
\begin{center}
\begin{tabular}{cc}
\includegraphics[scale=0.5]{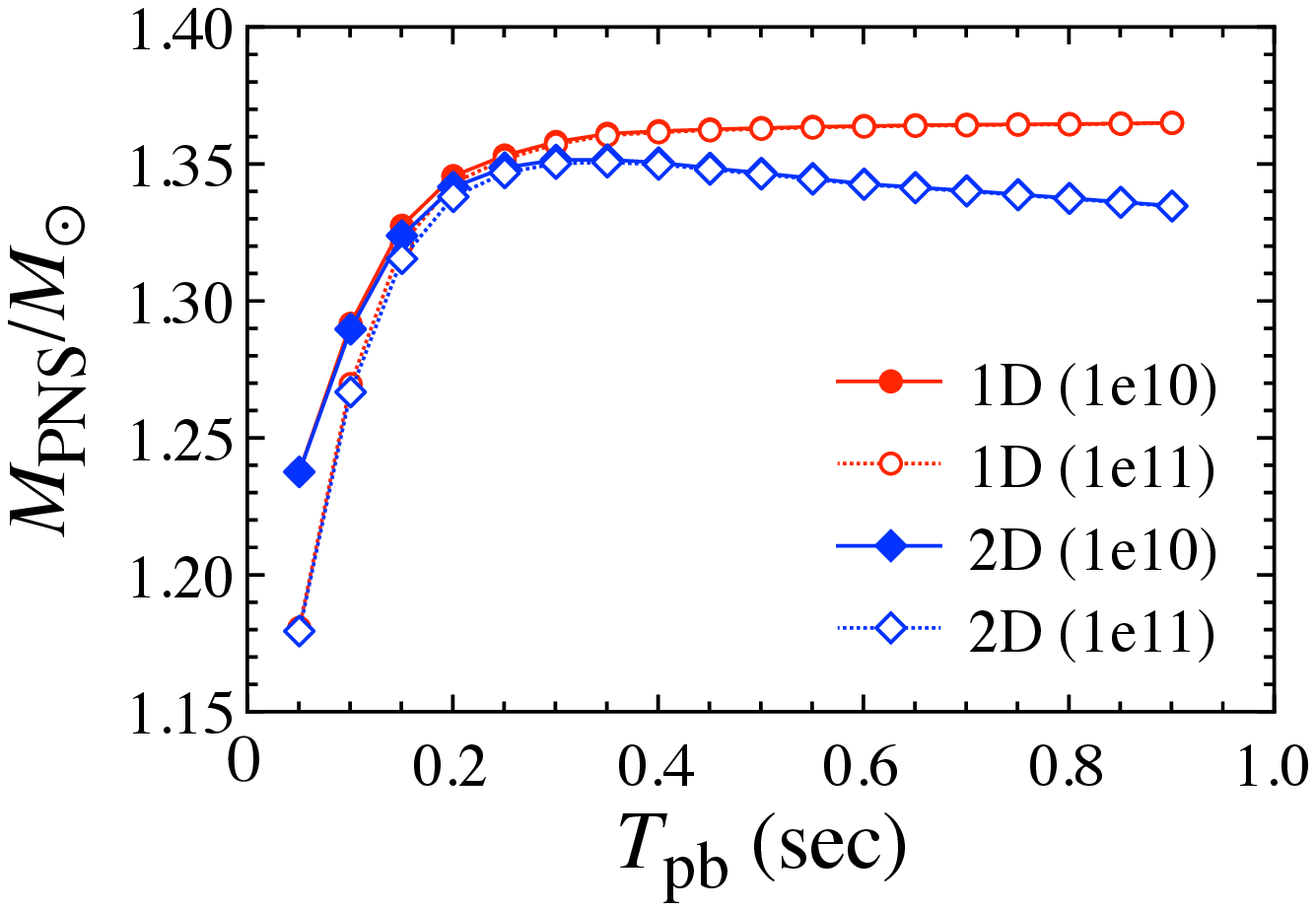} &
\includegraphics[scale=0.5]{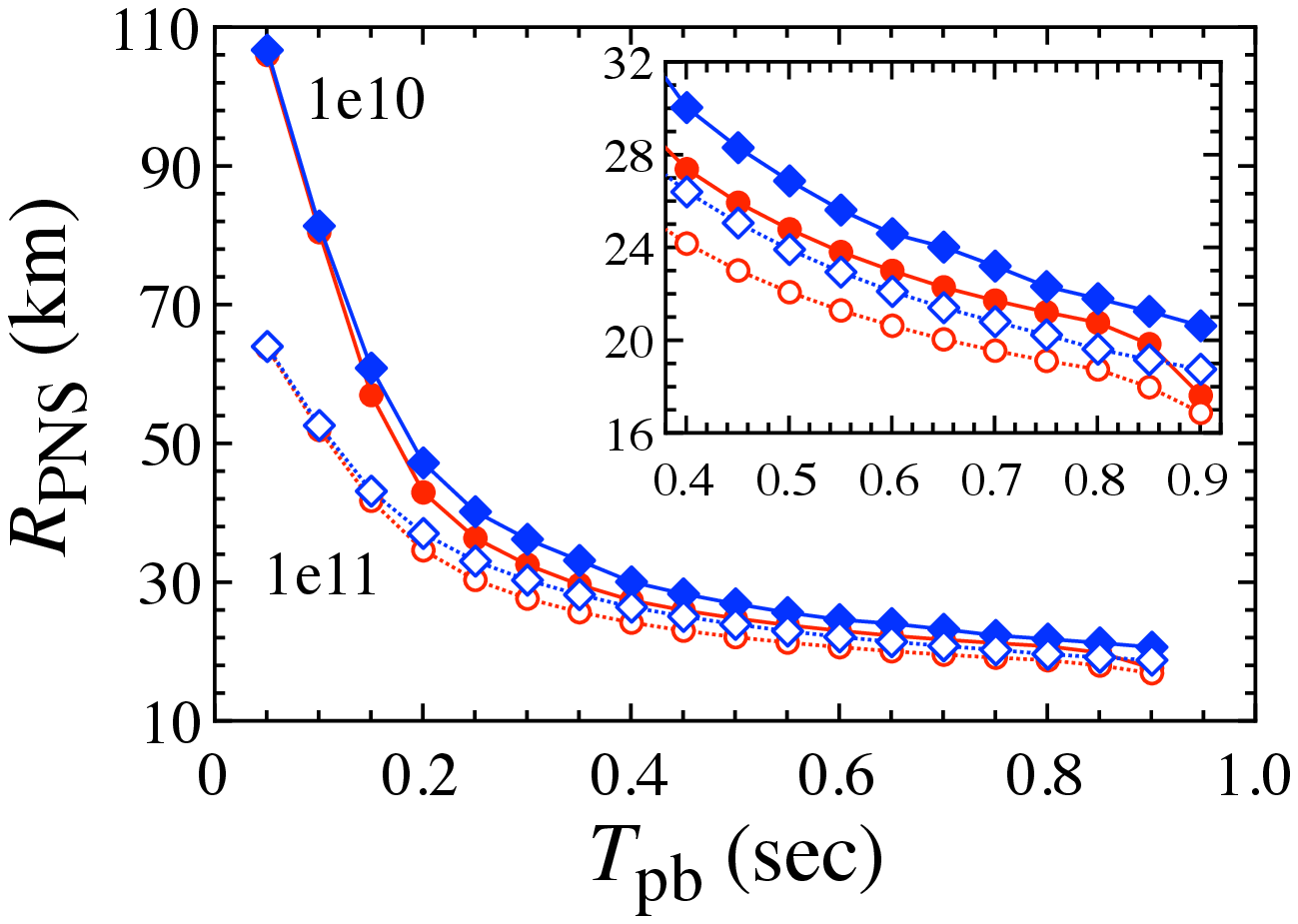}  
\end{tabular}
\end{center}
\caption{%%
Time evolution of PNS mass (left) and radius (right), where $T_{\rm pb}$ is the time after core bounce. The circles and diamonds correspond to the results for the PNS models obtained from 1D and 2D simulation, respectively. The surface density is selected as $\rho_s=10^{11}$ g/cm$^3$ (open marks) and $10^{10}$ g/cm$^3$ (filled marks). Since the 2D model explodes at 280 ms, the matter at the surface of PNS is ejected after that. Then $M_{\rm PNS}$ is decreased as a function of time. On the other hand, 1D model cannot explode and the mass accretion to the PNS continues. In this model, the mass accretion rate is low and $M_{\rm PNS}$ becomes almost constant after 300 ms. 
}%%
\label{fig:MRt}
\end{figure*}

The PNS models are prepared by averaging the properties in the angular direction for the 2D numerical simulation\footnote{If one considers the PNS asteroseismology with numerical simulations with rotating progenitor models and the rotational effects are significantly important in the PNS evolution, or if the multi-dimensional effects, such as convective motions, become significantly inside the PNSs, the procedure for spherically averaging may strongly constrain the PNS models, where some effects may forcedly be killed out.}. We remark that the turbulent motions are violent in the early phase after core bounce (so-called prompt convection), with which the PNS models may not be an appropriate as a spherically symmetric background, but we simply consider the PNS models at each 50 ms after core bounce as in the previous studies. Then, the PNS surface is determined with a specific surface density, $\rho_s$. In Fig. \ref{fig:MRt}, we show the evolution of PNS mass ($M_{\rm PNS}$) and radius ($R_{\rm PNS}$) with time after core bounce ($T_{\rm pb}$), where the circles and diamonds denote the results obtained with 1D and 2D simulations, respectively, while open and filled marks correspond to the PNS models with $\rho_s=10^{11}$ and $10^{10}$ g/cm$^3$. Inside the right panel of Fig. \ref{fig:MRt}, we also show the enlarged view. From this figure, one can see that the PNS mass is almost independent of the selection of $\rho_s$ after $T_{\rm pb}\sim 0.3$ sec, where the PNS mass obtained in 2D simulations decreases with time because the progenitor mass is so small that the explosion is done well. We also find that the dependence of the PNS radius on $\rho_s$ becomes weaker with time, i.e., $(R_{\rm PNS}^{10}-R_{\rm PNS}^{11})/R_{\rm PNS}^{11}=54.7 \%$, $19.5 \%$, $11.3 \%$, and $10.0 \%$ at $T_{\rm pb}= 0.1, 0.3, 0.6$, and 0.9 sec, where $R_{\rm PNS}^{10}$ and $R_{\rm PNS}^{11}$ respectively denote the PNS radius with $\rho_s=10^{10}$ and $10^{11}$ g/cm$^3$ obtained from 2D simulations. 

In this study, we mainly discuss by using the PNS model constructed from the 2D simulation with He2.9 and LS220, but for reference we also consider the other PNS models. So, in Table \ref{tab:PNS-model} we list the PNS models discussed in this article, where the label of the PNS model is named by the EOS and dimension of the numerical simulation, e.g., LS220-2D for the PNS model constructed with LS220 by the 2D simulation.

%%%%%%%%%%%%%%%%%%%%%%%%%%%%%%%%%%%
% Figure 1
%%%%%%%%%%%%%%%%%%%%%%%%%%%%%%%%%%%
%%%%%%%%%%%%%%%%%%%%%%%%%%%%%%%%%%%

%%%%%%%%%%%%%%%%%%%%%%%%%%%%%%%%%%%
% Table 1
%%%%%%%%%%%%%%%%%%%%%%%%%%%%%%%%%%%
\begin{table}
\centering
\caption{PNS models discussed in this study. For each PNS model, we list the name of PNS model, the progenitor model, EOS, the dimension of numerical simulation, the position in text where we discuss, and the reference for the linear analysis on each PNS model. 
}
%\begin{tabular}{cccc}
%\hline\hline
%  label & progenitor model & EOS & dimension  \\
%\hline
% LS220-2D & $2.9M_\odot$ \citep{He29} & LS220 \citep{LS220}  & 2D \citep{takiwaki2020usm}  \\
% LS220-1D &                          &                       & 1D \citep{takiwaki2020usm}  \\
% SFHx-3D  & $15M_\odot$ \citep{WW95}  & SFHx \citep{SHF2013}   & 3D \citep{KKT2016}  \\ 
% TGTF-2D  & $20M_\odot$ \citep{WH07}  & TGTF \citep{Togashi17} & 2D \citep{takiwaki2020eos}  \\
% DD2-2D   & $20M_\odot$ \citep{WH07}  & DD2 \citep{DD2}        & 2D \citep{takiwaki2020eos}  \\
%\hline\hline
%\end{tabular}
\begin{tabular}{cccccc}
\hline\hline
  label & progenitor model & EOS & dimension & corresponding portion & linear analysis \\
\hline
 LS220-2D & $2.9M_\odot$$^a$ & LS220$^d$  & 2D$^h$ & Sec. \ref{sec:Oscillation}    & this study \\
 LS220-1D &                          &                       & 1D$^h$ & Appendix \ref{sec:appendix_2} & this study \\
 SFHx-3D  & $15M_\odot$$^b$  & SFHx$^e$  & 3D$^i$  & Sec. \ref{sec:Oscillation}  & \cite{SKTK2017,SKTK2019} \\ 
 TGTF-2D  & $20M_\odot$$^c$  & TGTF$^f$ & 2D$^j$ & Appendix \ref{sec:appendix_1} & \cite{ST2020} \\
 DD2-2D   & $20M_\odot$$^c$  & DD2$^g$   & 2D$^j$ & Appendix \ref{sec:appendix_1} & \cite{ST2020} \\
\hline\hline
\multicolumn{6}{l}{$^a$\cite{He29}, $^b$\cite{WW95}, $^c$\cite{WH07}.}\\
\multicolumn{6}{l}{$^d$\cite{LS220}, $^e$\cite{SHF2013}, $^f$\cite{Togashi17}, $^g$\cite{DD2}.}\\
\multicolumn{6}{l}{$^h$\cite{takiwaki2020usm}, $^i$\cite{KKT2016}, $^j$\cite{takiwaki2020eos}.}\\
%Notes. $^a$\cite{He29}, $^b$\cite{WW95}, $^c$\cite{WH07}, $^d$\cite{LS220}, $^e$\cite{SHF2013}, $^f$\cite{Togashi17}, $^g$\cite{DD2}, 
\end{tabular} 
\label{tab:PNS-model}
\end{table}
%%%%%%%%%%%%%%%%%%%%%%%%%%%%%%%%%%%

%%%%%%%%%%%%%%%%%%%%%%%%%%%%%%%%%%%
% Table 2
%%%%%%%%%%%%%%%%%%%%%%%%%%%%%%%%%%%
%\begin{table}
%\centering
%\caption{PNS models discussed in this study. For each PNS model, we list the name of PNS model, the position in text where we discuss, and the reference for the linear analysis on each PNS model.
%}
%\begin{tabular}{ccc}
%\hline\hline
%   & corresponding portion & linear analysis    \\
%\hline
% LS220-2D  & Sec. \ref{sec:Oscillation}          & this study     \\
% LS220-1D  & Appendix \ref{sec:appendix_2} & this study     \\
% SFHx-3D  & Sec. \ref{sec:Oscillation}  & \cite{SKTK2017,SKTK2019}     \\ 
% TGTF-2D  & Appendix \ref{sec:appendix_1} & \cite{ST2020}    \\
% DD2-2D  & Appendix \ref{sec:appendix_1} & \cite{ST2020}        \\
%\hline\hline
%\end{tabular}
%\label{tab:PNS-model2}
%\end{table}
%%%%%%%%%%%%%%%%%%%%%%%%%%%%%%%%%%%

%%%%%%%%%%%%%%%%%%%%%%%%%%%%%%%%%%%%%%%%%%%%%%%%
\section{Gravitational wave signals from PNS}
\label{sec:Oscillation}
%%%%%%%%%%%%%%%%%%%%%%%%%%%%%%%%%%%%%%%%%%%%%%%%

On the PNS models obtained via 2D simulation, we make a linear analysis. For this purpose, as in \cite{ST2016,SKTK2019,SS2019,ST2020}, we simply adopt the relativistic Cowling approximation in this study, where the metric perturbation is neglected during the fluid oscillations. In this case, the perturbation equations can be derived by linearizing the energy-momentum conservation law. In addition, one has to impose appropriate boundary conditions at the stellar center and the outer boundary, i.e., the PNS surface. The concrete perturbation equations and the boundary conditions are the same as in \cite{SKTK2019}. Then, the problem to solve becomes an eigenvalue problem with respect to the eigenvalue, $\omega$, with which the eigenfrequency, $f$, is determined via $f=\omega/(2\pi)$. As the standard standard asteroseismology, the eigenmodes are identified by counting the nodal numbers in the eigenfunctions, i.e., the nodal numbers of $f$-, pressure ($p_i$-), and $g_i$-modes are 0, $i$, and $i$, respectively. With respect to some of eigenmodes (especially $p_i$- and $g_i$-modes with lower $i$ and $f$-mode) in early phase after core bounce, the nodal numbers become more than their definition because the additional nodes appear in the vicinity of the stellar center. Even in such a case, the nodal numbers for the $p_i$- and $g_i$-modes with higher $i$, e.g., $i\gsim 3$, are the same as the definition. So, even for the eigenmodes whose nodal numbers are more than their definition, we simply classify them as usual by using the $p_i$- and $g_i$-modes with higher $i$. 
%we can identified the eigenmodes with checking the shape of the eigenfunction. 

%%%%%%%%%%%%%%%%%%%%%%%%%%%%%%%%%%%
% Figure 2
%%%%%%%%%%%%%%%%%%%%%%%%%%%%%%%%%%%
\begin{figure}
\begin{center}
\includegraphics[scale=0.6]{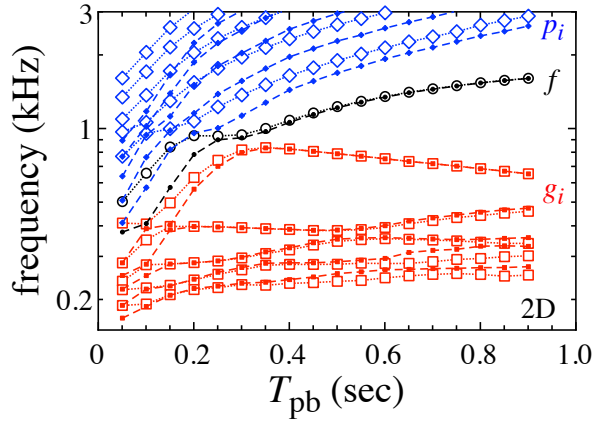}  
\end{center}
\caption{%%
Comparing the $f$-, $p_i$-, and $g_i$-mode frequencies for $i=1$ up to $5$ on the PNS models with $\rho_s=10^{11}$ g/cm$^3$ to those with $\rho_s=10^{10}$ g/cm$^3$, where the open marks with dotted lines correspond to the results with $\rho_s=10^{11}$ g/cm$^3$, while the filled marks with dashed lines are the results with $\rho_s=10^{10}$ g/cm$^3$.
}%%
\label{fig:ft-2D}
\end{figure}
%%%%%%%%%%%%%%%%%%%%%%%%%%%%%%%%%%%

First, in order to see how the time evolution of the eigenfrequencies of gravitational waves depends on the selection of $\rho_s$, in Fig. \ref{fig:ft-2D} we show the frequency evolution for the PNS models with $\rho_s=10^{11}$ g/cm$^3$ (open marks with dotted lines) and with $\rho_s=10^{10}$ g/cm$^3$ (filled marks with dashed lines), provided from the 2D simulation, where the circles, diamonds, and squares denote the $f$-, $p_i$-, and $g_i$-modes for $i=1$ up to 5. We remark that the frequencies increase (decrease) for $p_i$-modes ($g_i$-modes) as $i$ increases. From this figure, one can clearly observe a phenomenon of the avoided crossing in the time evolution of eigenfrequencies, as in \cite{MRBV2018,SS2019,ST2020,TCPOF2019a}. That is, for example one can see such a phenomenon between the $f$- and $g_1$-modes at $T_{\rm pb}\simeq 0.3$ sec.

\begin{figure}
\begin{center}
\includegraphics[scale=0.45]{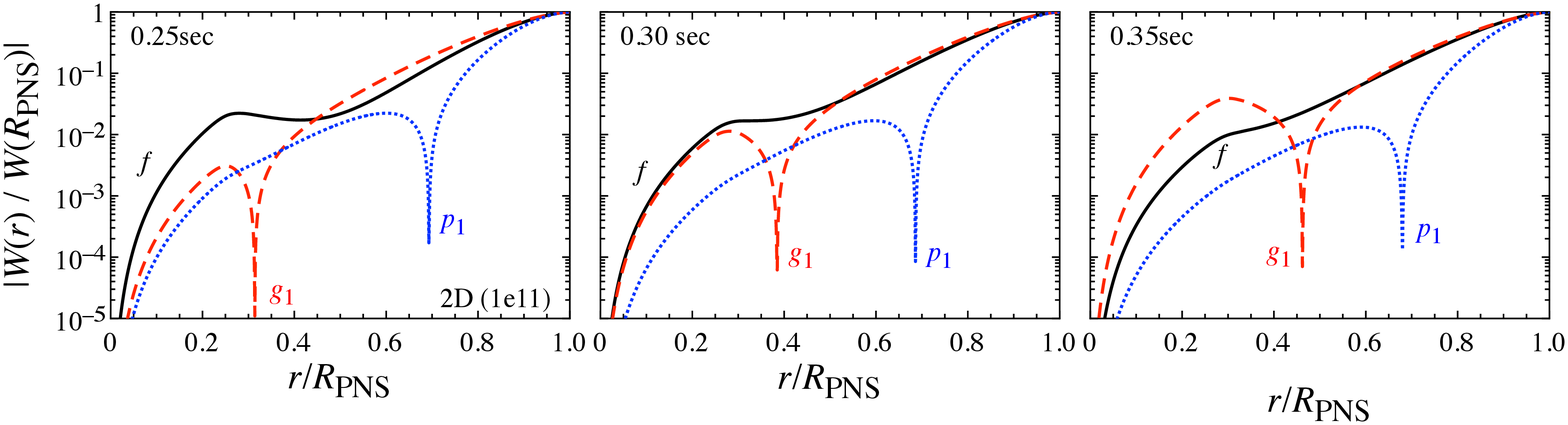}  
\end{center}
\caption{%%
Radial profile of the absolute value of the eigenfunction for the PNS model with $\rho_s=10^{11}$ g/cm$^3$ obtained from 2D simulations, where $W(r)$ denotes the eigenfunction of the radial displacement. The left, middle, and right panels correspond to the PNS models at $T_{\rm pb}=0.25$, $0.30$, and $0.35$ sec, respectively, where the solid, dashed, and dotted lines denote the eigenfunctions for the $f$-, $g_1$- and $p_1$-modes.
}%%
\label{fig:Wr_AC}
\end{figure}

In order to see the phenomena around the avoided crossing, in Fig. \ref{fig:Wr_AC} we show the radial profile of the absolute value of the eigenfunctions (the Lagrangian displacement in the radial direction) for the $f$-, $g_1$-, and $p_1$-modes, where the left, middle, and right panels correspond to the PNS models at $\simeq 0.25$, $0.30$, and $0.35$ sec. From this figure one can see that the amplitude of $g_1$-mode increases with time in the deeper region of the PNS, while the eigenfunction of the $p_1$-mode is almost unchanged during the avoided crossing between the $f$- and $g_1$-modes. We remark that the shape of the $f$-mode at $0.25$ sec is not as usual, comparing to that for the cold neutron stars, i.e., the amplitude of the $f$-mode at $0.25$ sec does not monotonically increase from the center to the surface. The behavior of the $f$- and $g_1$-modes seems to be consistent with the result shown in Fig. 5 in \cite{TCPOF2019a}. In addition, one can see that at least the eigenfunctions of the $g_1$-mode becomes very similar to that of the $f$-mode at the avoided crossing. We remark that the avoid crossing does not happen with the mode classification newly defined in  \cite{TCPF2018,TCPOF2019a}. 

%%%%%%%%%%%%%%%%%%%%%%%%%%%%%%%%%%%
% Figure 3
%%%%%%%%%%%%%%%%%%%%%%%%%%%%%%%%%%%
%%%%%%%%%%%%%%%%%%%%%%%%%%%%%%%%%%%

\begin{figure*}
\begin{center}
\begin{tabular}{cc}
\includegraphics[scale=0.5]{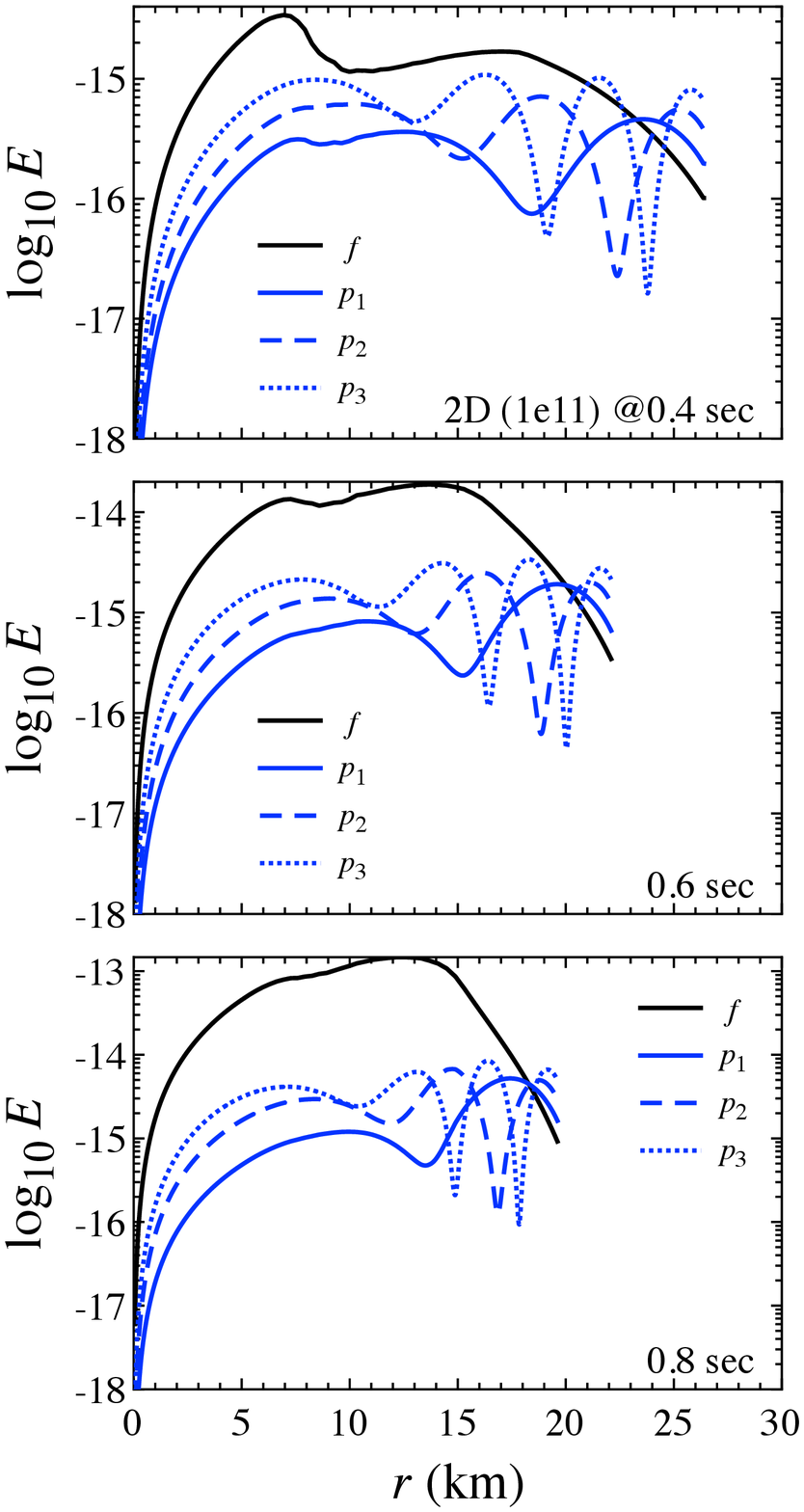} &  
\includegraphics[scale=0.5]{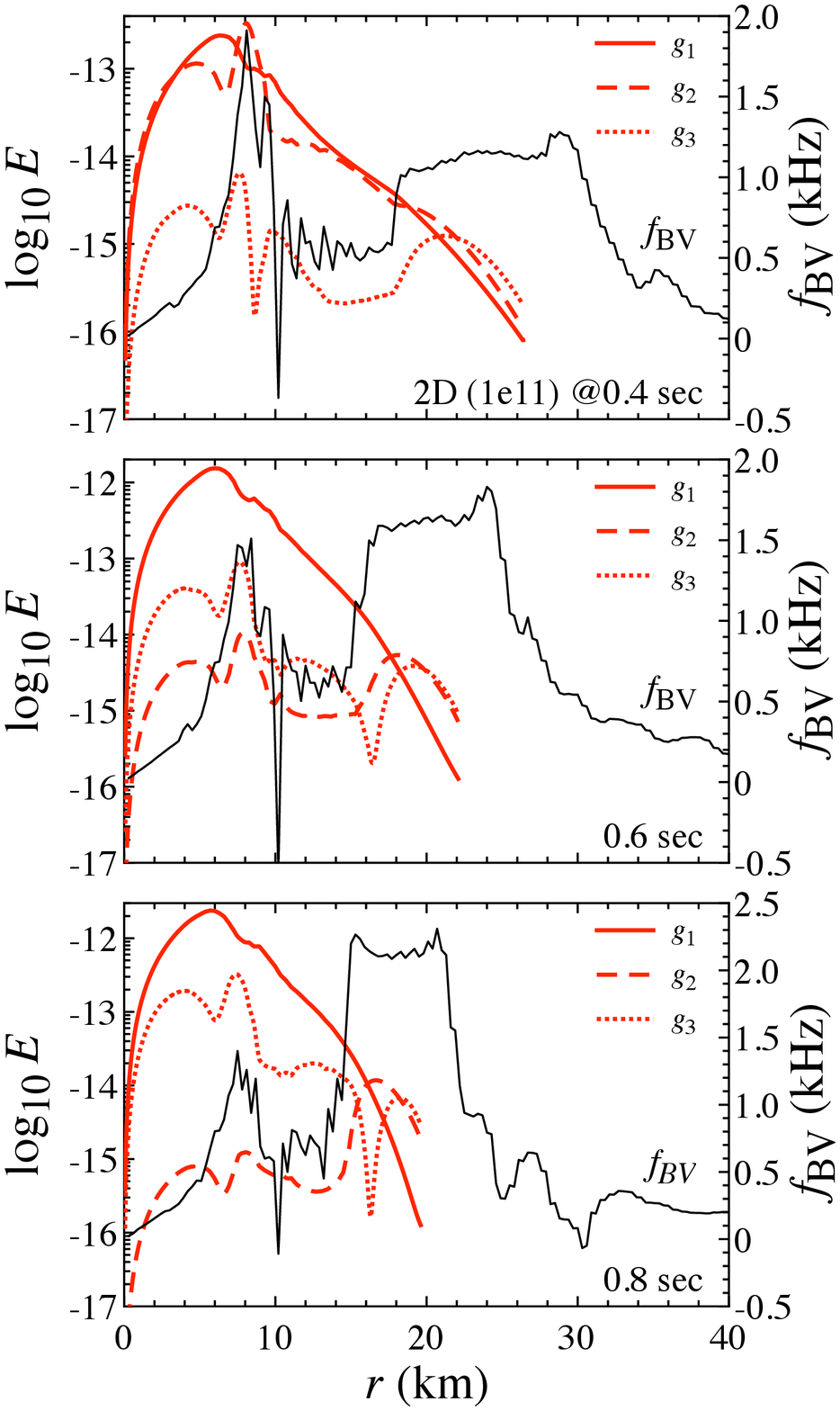}  
\end{tabular}
\end{center}
\caption{%%
Radial-dependent pulsation energy density, $E$, is shown for the $f$- and $p_i$-modes in the left panel and for the $g_i$-modes in the right panel, where the top, middle, and bottom panels correspond to the PNS models at $T_{\rm pb}\simeq 0.4$, $0.6$, and $0.8$ sec. In the right panel, for reference the Brunt-V\"{a}is\"{a}l\"{a} frequency, $f_{\rm BV}$ is also shown. The right endpoint of $E$ corresponds to the PNS surface for the corresponding time. We note that the $g_1$-mode frequencies at $0.4$, $0.6$, and $0.8$ sec after core bounce are respectively $825.0$, $753.0$, and $685.4$ Hz.
}%%
\label{fig:Er}
\end{figure*}

From Fig. \ref{fig:ft-2D}, we also find that the eigenfrequencies strongly depend on the selection of $\rho_s$ especially in the early phase after core bounce, such as until $T_{\rm pb}\sim 0.3$ sec, while we also confirm that the $f$- and $g_1$-modes are independent of $\rho_s$ except for the early phase. This result is more or less consistent with that shown in  \cite{MRBV2018}, which may come from a behavior of the pulsation energy density, $E$, corresponding eigenfunctions. Here, the Newtonian radial-dependent energy density is estimated as in \cite{MRBV2018,SKTK2019,TCPF2018}, i.e., 
\begin{equation}
  E(r) \sim \frac{\omega^2\varepsilon}{r^4}\left[W^2 + \ell(\ell+1)r^2V^2\right],
\end{equation}
where $\varepsilon$, $\omega$, and $V$ are the energy density, the eigenvalue, and the Lagrangian displacement in the angular direction. As an example, in Fig. \ref{fig:Er} we show $E(r)$ for the $f$- and $p_i$-modes in the left panel and for the $g_i$-modes in the right panel, where the top, middle, and bottom panels correspond to the PNS models at $T_{\rm pb}\simeq 0.4$, $0.6$, and $0.8$ sec. From this figure, one can see that the $f$- and $g_1$-modes strongly oscillate inside the PNS. On the other hand, the other modes strongly oscillate not only inside the PNS but also the surface region of PNS. In addition, the position of node for these modes (except for the $f$- and $g_1$-modes) exist closer to the surface. This may be a reason why the $f$- and $g_1$-modes are less sensitive to the position of the PNS surface (or the selection of $\rho_s$).
We should also mention the discrepancy between the current results and our previous results in \cite{SKTK2019}, where the frequencies strongly depend on the selection of $\rho_{\rm s}$. This is because the PNS models considered in  \cite{SKTK2019} are quite unusual, where the standing accretion-shock instability (SASI) is so strong that almost whole region inside the PNS is convectively blended. As a result, almost whole region inside the PNS becomes convectively unstable, as shown in Fig. 3  in \cite{SKTK2019}. On the other hand, with using the usual PNS models as in this study, we can show that the $f$- and $g_1$-mode frequencies depend weakly on the selection of $\rho_{\rm s}$. 

Moreover, in the right panel of Fig. \ref{fig:Er}, we also show the Brunt-V\"{a}is\"{a}l\"{a} frequency, $f_{\rm BV}$, for reference. We remark that $f_{\rm BV}$ is determined via only the background (unperturbed) properties as
\begin{equation}
 f_{\rm BV} = {\rm sgn}({\cal N}^2)\sqrt{|{\cal N}^2|/2\pi},
\end{equation}
where ${\cal N}^2$ is given by
\begin{equation}
 {\cal N}^2 = -e^{2\Phi-2\Lambda}\frac{\Phi'}{\varepsilon + p}\left(\varepsilon' - \frac{p'}{c_s^2}\right).
\end{equation}
In this equation, $\Phi$ and $\Lambda$ are the metric function as $g_{tt}=-e^{2\Phi}$ and $g_{rr}=e^{2\Lambda}$, $p$ and $c_s$ denote the pressure and sound velocity, and the prime denotes the partial derivative with respective to $r$.  We remark that the region with ${\cal N}^2>0$ (${\cal N}^2<0$) is convectively stable (unstable) region \footnote{The statement about the stability mentioned in \cite{SKTK2019} is not correct, where the condition is opposite to what they mentioned. That is, the region with ${\cal A} < 0$ (or ${\cal N}^2 > 0$) is a stable region. So, most of the PNS region is convectively unstable for the models discussed in \cite{SKTK2019}, which may be a reason why the $g_i$-modes could not be found in \cite{SKTK2019}.
On the other hand, since the region, where the Brunt-V\"{a}is\"{a}l\"{a} frequency becomes negative, i.e., convectively unstable, is very limited in this study, the $g$-mode oscillations are stably excited. Thus, whether or not the $g$-mode oscillations can be excited strongly depends on the strength of convection and the width of the convectively unstable region.}. From this figure, one can see the peak in $f_{\rm BV}$ appears at $\sim 8$ km and in the vicinity of the PNS surface, where the peak at $\sim 8$ km decreases and that in the vicinity of the PNS surface increases with time. The decrease of $f_{\rm BV}$ around 8 km may  correspond to the decrease of the $g_1$-mode frequency with time, i.e., the $g_1$-mode frequencies at $\sim 0.4$, $0.6$, and $0.8$ sec are respectively $825.0$, $753.0$, and $685.4$ Hz. In addition, we find that the shape of pulsation energy density for $g_2$ and $g_3$-modes strongly depend on the $f_{\rm BV}$ distribution. In particular, the pulsation energy of the $g_2$-mode becomes more dominant in the vicinity of the PNS surface with time  due to the enhancement of $f_{\rm BV}$ in the region around the PNS surface. On the other hand, the pulsation energy of the $g_3$-mode is still stronger in the core region of PNS. So, we may say that the $g_1$- and $g_3$-modes correspond to the core $g$-mode, while the $g_2$-mode is the surface $g$-mode.

%%%%%%%%%%%%%%%%%%%%%%%%%%%%%%%%%%%
% Figure 4
%%%%%%%%%%%%%%%%%%%%%%%%%%%%%%%%%%%
%%%%%%%%%%%%%%%%%%%%%%%%%%%%%%%%%%%

\begin{figure}
\begin{center}
\includegraphics[scale=0.6]{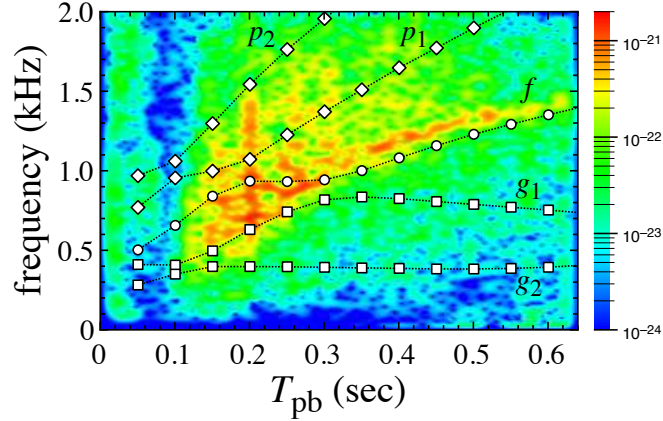}  
\end{center}
\caption{%%
Comparison between the gravitational wave signals obtained from the numerical simulation (background contour) and several eigenfrequencies for the PNS with $\rho_s=10^{11}$ g/cm$^3$, where circles, diamonds, and squares denote the $f$-, $p_i$-, and $g_i$-modes for $i=1$ or 2. The source distance is assumed to be $D=10$ kpc. 
}%%
\label{fig:2D-spe}
\end{figure}

Next, we consider to identify the ramp up signals of gravitational waves in numerical data. Using the numerical data obtained via hydrodynamical simulations, as in \cite{Murphy09}, the dimensionless characteristic gravitational wave strain is given by
\begin{equation}
  h_{\rm char}(f,T_{\rm pb}) = \sqrt{\frac{2G}{\pi^2 c^3 D^2}\frac{dE_{\rm GW}}{df}}, \label{eq:strain}
\end{equation}
where $D$ denotes the source distance, while $dE_{\rm GW}/df$ denotes the time-integrated energy spectra of gravitational wave calculated with a short-time Fourier transform, $\tilde{S}(f,T_{\rm pb})$, via
\begin{eqnarray}
  \frac{dE_{\rm GW}}{df}(f,T_{\rm pb}) &=& \frac{3G}{5c^2}\left(2\pi f\right)^2|\tilde{S}(f,T_{\rm pb})|, \\
  \tilde{S}(f,T_{\rm pb}) &=& \frac{1}{2}\int_{T_{\rm pb}-\Delta t}^{T_{\rm pb}+\Delta t} \frac{d^2\Ibar_{zz}}{dt^2}
       \left[1+\cos\left(\frac{\pi(t-T_{\rm pb})}{2\Delta t}\right)\right]\exp(-2\pi i ft)dt,   \label{eq:S}
\end{eqnarray}
where $2\Delta t$ denotes the width of the window function and $\Ibar_{zz}$ is the $zz$-component of the reduced mass-quadrupole tensor $\Ibar_{jk}$ given by Eq. (11) in \cite{Murphy09}. In Fig. \ref{fig:2D-spe}, we show the resultant value of $h_{\rm char}$ with contour, adopting that $D=10$ kpc and $\Delta t = 20$ ms. In this figure, one can clearly observe the ramp up signals from $\sim 500$ hertz up to $\sim 1.5$ kilohertz in the time interval of $T_{\rm pb}\simeq 0.15-0.65$ sec. On this figure, we also plot the several eigenfrequencies on PNS model with $\rho_s=10^{11}$ g/cm$^3$. From this figure, it is obviously found that the ramp up signals correspond well to the $g_1$-mode in the early phase and to the $f$-mode after the avoided crossing. But, since the $g_1$-mode frequency depends on $\rho_s$ in the early phase as mentioned before, it is not sure whether or not the ramp up signal corresponds well to the $g_1$-mode for different PNS models provided with the different numerical simulations. In order to check this point, we calculate the gravitational wave signals from the 2D numerical simulations with completely different progenitor models and EOSs as in Table \ref{tab:PNS-model} and compare it with the eigenmodes calculated for the corresponding PNS with $\rho_{\rm s}=10^{11}$ g/cm$^3$.  Then, we find that the ramp up signals still seem to be good agreement with the $g_1$-mode on the PNS model with $10^{11}$ g/cm$^3$ as shown in Fig. \ref{fig:2D-spect} (see the details in Appendix \ref{sec:appendix_1}).

%%%%%%%%%%%%%%%%%%%%%%%%%%%%%%%%%%%
% Figure 5
%%%%%%%%%%%%%%%%%%%%%%%%%%%%%%%%%%%
%%%%%%%%%%%%%%%%%%%%%%%%%%%%%%%%%%%

\begin{figure*}
\begin{center}
\begin{tabular}{cc}
\includegraphics[scale=0.5]{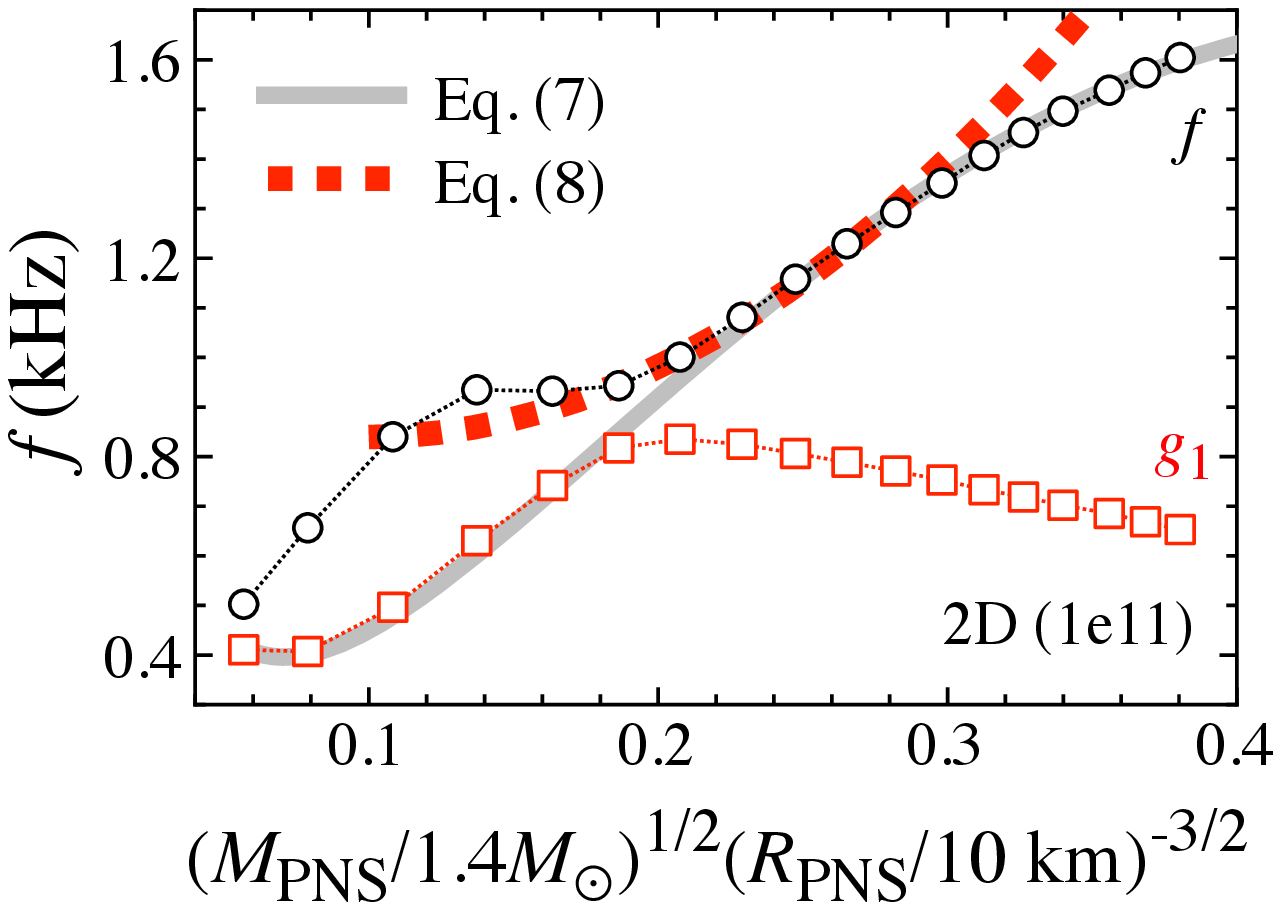} &  
\includegraphics[scale=0.5]{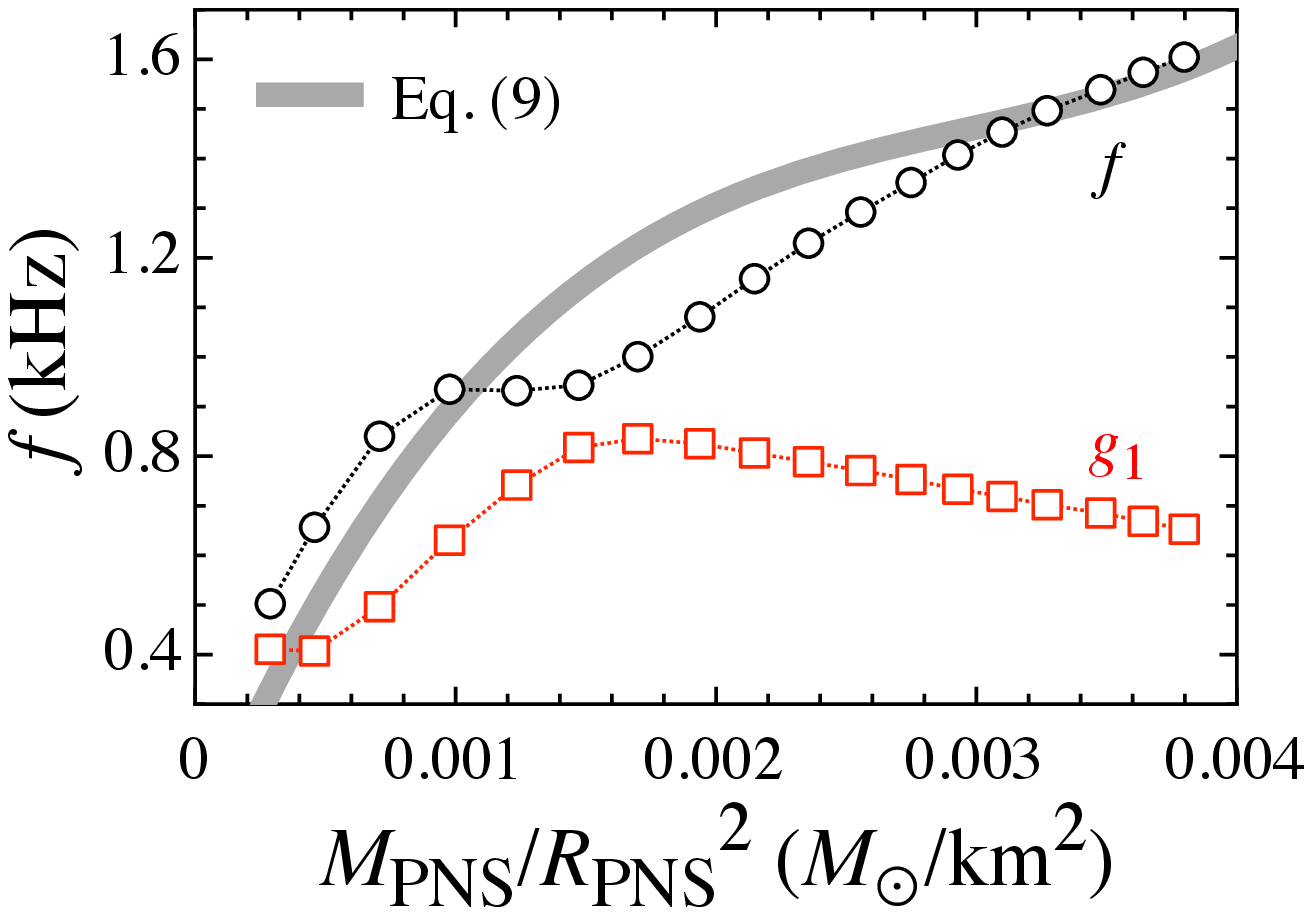}  
\end{tabular}
\end{center}
\caption{%%
In the left panel, frequencies of the $f$- and $g_1$-modes are shown as a function of the square root of the PNS average density, where the thick-solid and thick-dotted lines are the fitting formula given by Eqs. (\ref{eq:fit}) and (\ref{eq:SS19}). In the right panel, frequencies of the $f$- and $g_1$-modes are shown as a function of the surface gravity of the PNS, where the thick-solid line denote the fitting formula given by Eq. (\ref{eq:TF19}). 
}%%
\label{fig:ft-fit}
\end{figure*}

Now, it is observationally important what one can learn from the direct observation of the gravitational wave signals after supernova explosion, assuming that principal signals are the ramp up signals appearing in numerical simulations. That is, since the ramp up signals partially correspond to the $f$- and $g_1$-mode frequencies, it is very useful if one could connect these frequencies to the PNS properties. In the left panel of Fig. \ref{fig:ft-fit}, we show the $f$- and $g_1$-mode frequencies for the PNS model with $\rho_s=10^{11}$ g/cm$^3$ as a function of the square root of the normalized PNS average density, $(M_{\rm PNS}/1.4 M_\odot)^{1/2}(R_{\rm PNS}/10 {\rm km})^{-3/2}$. With this data, we successfully find that the $f$- and $g_1$-mode frequencies, which correspond to the ramp up signals, are well expressed as
\begin{equation}
  f {\rm (kHz)} = -3.250 -0.978\ln(x) + 15.984x -15.051x^2,  \label{eq:fit}
\end{equation}
where $x$ is the square root of the normalized PNS average density, i.e., $x=(M_{\rm PNS}/1.4 M_\odot)^{1/2}(R_{\rm PNS}/10 {\rm km})^{-3/2}$. In practice, the frequency predicted from Eq. (\ref{eq:fit}) is also plotted with the thick-solid line in the left panel of Fig. \ref{fig:ft-fit}. Thus, using Eq. (\ref{eq:fit}), one could get the evolution of the PNS average density via the observed frequency of gravitational wave after supernova explosion. In this study, since we consider only one progenitor model and one EOS, it is difficult to say how this relation is independent of the models, i.e., Eq. (\ref{eq:fit}) can not be used to analyse data from future gravitational wave observations. Even so, this relation seems to be independent of the models at least in the early phase, as shown in Fig. \ref{fig:ft-fita} in Appendix \ref{sec:appendix_1}. Anyway, additional models should be considered in the future.

The relation similar to Eq. (\ref{eq:fit}) has already been proposed, as a function of $x$ in \cite{SS2019};
\begin{equation}
   f {\rm (kHz)} = 0.9733 - 2.7171x + 13.7809x^2, \label{eq:SS19}
\end{equation}
and as a function of $\bar{x}\equiv M_{\rm PNS}/R_{\rm PNS}^2$ in the unit of $M_\odot$/km$^2$ in \cite{TCPOF2019b};
\begin{equation}
   f {\rm (kHz)} = 12.4 \times 10^{2} \bar{x} - 378 \times 10^{3}\bar{x}^2 + 4.24\times 10^{7} \bar{x}^3, \label{eq:TF19}
\end{equation}
although in \cite{TCPOF2019b} the ramp up signal is identified as $g_2$-mode in their classification. Eq. (\ref{eq:SS19}) are derived for the $f$-mode frequency after the avoided crossing with the $g_1$-mode with the PNS models provided by the 1D numerical simulations, which are eventually collapsed into black hole. In the left panel of Fig. \ref{fig:ft-fit}, we also plot the thick-dotted line given by Eq. (\ref{eq:SS19}). From this figure, we find that the $f$-mode frequency is well identified by Eq. (\ref{eq:SS19}) up to $f\simeq 1.4$ kHz, which corresponds to $T_{\rm pb}\lsim 0.6$ sec from Fig. \ref{fig:2D-spe} (or Fig. \ref{fig:ft-2D}), but it deviates for $T_{\rm pb}\gsim 0.6$ sec. This may come from that  the PNS models for deriving Eq. (\ref{eq:SS19}) become more compact in the later phase due to the massive progenitor model. On the other hand, in order to see the correspondence between the eigenmodes calculated in this study and Eq. (\ref{eq:TF19}), we show the $f$- and $g_1$-mode frequencies as a function of $\bar{x}$ in the right panel of Fig. \ref{fig:ft-fit}. This fitting formula seems to  correspond to the $f$-mode rather than the $g_1$-mode in the very early phase and also in the late phase, at least for comparing to our results.

%%%%%%%%%%%%%%%%%%%%%%%%%%%%%%%%%%%
% Figure 6
%%%%%%%%%%%%%%%%%%%%%%%%%%%%%%%%%%%
%%%%%%%%%%%%%%%%%%%%%%%%%%%%%%%%%%%

\begin{figure*}
\begin{center}
\begin{tabular}{cc}
\includegraphics[scale=0.5]{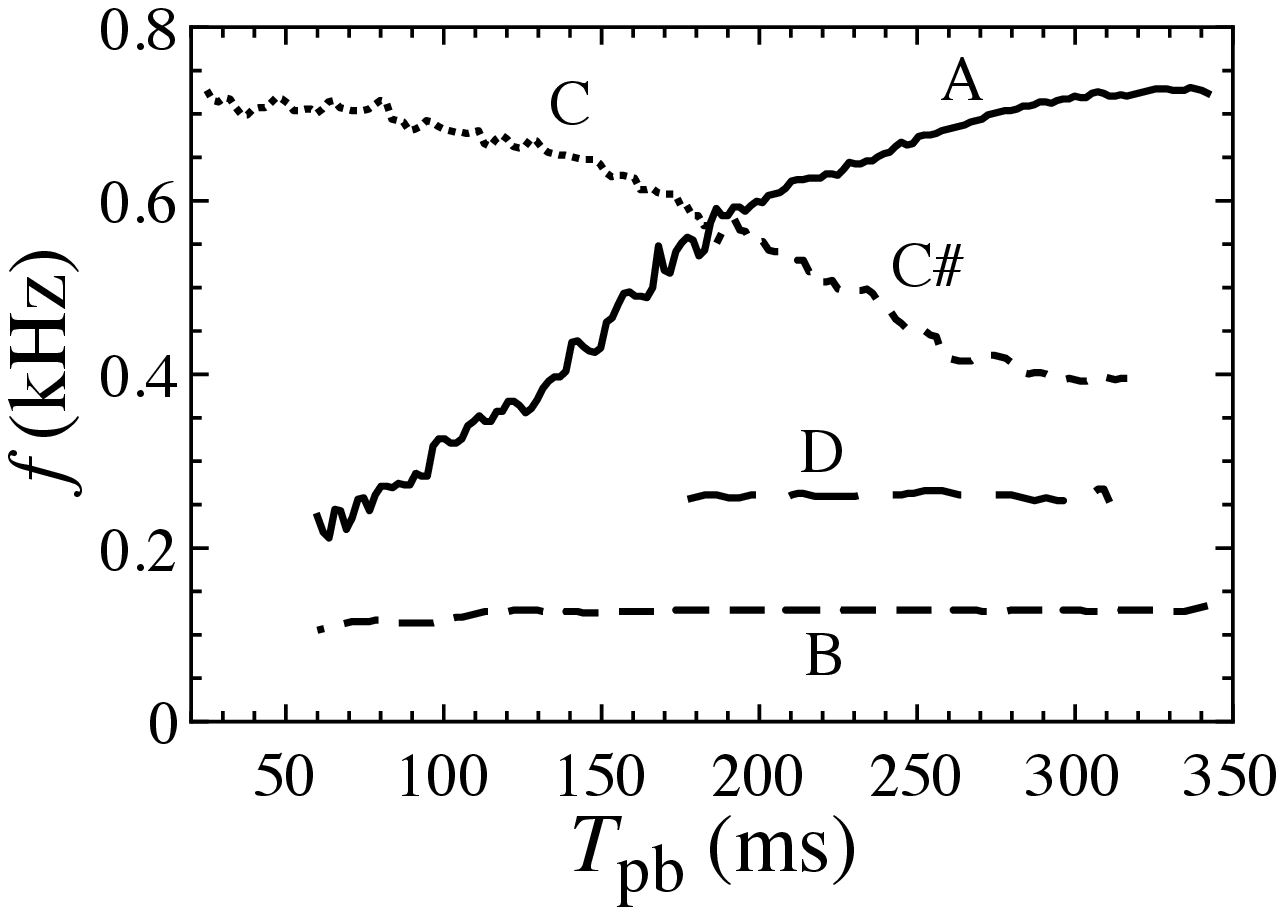} &
\includegraphics[scale=0.5]{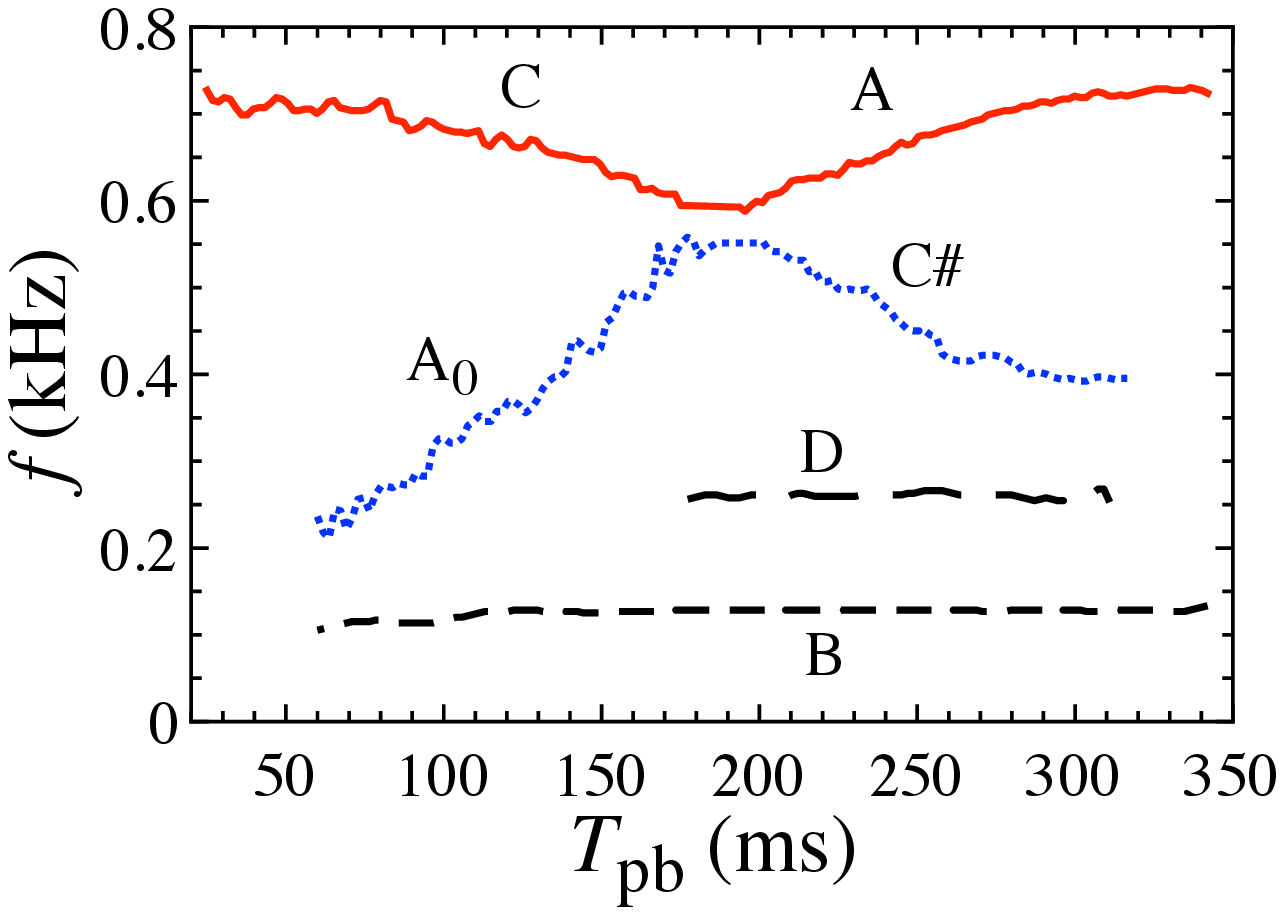}  
\end{tabular}
\end{center}
\caption{%%
Characteristic gravitational wave frequencies extracted by the time-frequency analysis \citep{Kawahara18} from the general relativistic 3D numerical simulation with SFHx \citep{KKT2016} in the left panel. The characteristic gravitational frequencies speculated with the avoided crossing, using the result shown in the left panel, in the right panel.
}%%
\label{fig:Kawa}
\end{figure*}

Finally, we make a comment with respect to the previous results about the gravitational wave signals, especially obtained via the general relativistic 3D simulation \citep{KKT2016}, which is done with a $15M_\odot$ progenitor model \citep{WW95} and SFHx EOS \citep{SHF2013}. Unlike most of the other simulation results, several modes of gravitational wave signals have been found in this simulation. The characteristic gravitational wave frequencies extracted by the time-frequency analysis \citep{Kawahara18} are shown in the left panel of Fig. \ref{fig:Kawa}, where the signal A corresponds to the ramp up signals. Nevertheless, considering the results with the PNS asteroseismology shown in Fig. \ref{fig:ft-2D}, the mode crossing between the signal A and C-C\# should be avoided, i.e., the gravitational wave signals may theoretically become as in the right panel of Fig. \ref{fig:Kawa}. If so, the signals A$_0$-C\# and C-A may correspond to the $g_1$- and $f$-modes, respectively, comparing to the result shown in Fig. \ref{fig:ft-2D}. In addition, the signal C may come from the $p_2$-, $p_1$-, and $f$-modes, which are exchanged through the avoided crossing, although the corresponding avoided crossing can not be seen obviously in the simulation data. On the other hand, the signals D and B, which are considered as a results of the SASI, may correspond to some of $g_i$-modes, e.g., the signal D may correspond to the $g_2$-mode. The correspondence mentioned here is just a speculation, but it would be confirmed in the future via more complicated analysis, e.g., with which one can distinguish the left and right panels in Fig. \ref{fig:Kawa}.

%%%%%%%%%%%%%%%%%%%%%%%%%%%%%%%%%%%
% Figure 7
%%%%%%%%%%%%%%%%%%%%%%%%%%%%%%%%%%%
%%%%%%%%%%%%%%%%%%%%%%%%%%%%%%%%%%%

%%%%%%%%%%%%%%%%%%%%%%%%%%%%%%%%%%%%%%%%%%%%%%%%
\section{Conclusion}
\label{sec:Conclusion}
%%%%%%%%%%%%%%%%%%%%%%%%%%%%%%%%%%%%%%%%%%%%%%%%

In order to understand the ramp up signals of gravitational waves appearing in the numerical simulations, we made a linear perturbation analysis by solving the eigenvalue problem on the PNS models, which are produced by the 2D numerical simulation with $2.9M_\odot$ He star as a progenitor model and with LS220 EOS. We found that the ramp up signals corresponds well to the $g_1$-mode in the early phase and to the $f$-mode after avoided crossing between the $f$- and $g_1$-modes of the PNS model. The results are basically consistent with the previous work \citep{MRBV2018,SS2019,ST2020,TCPOF2019a}. In addition, we successfully found the fitting formula for the $g_1$- and $f$-mode frequencies, which correspond to the ramp up signals, as a function of the PNS average density. Thus, assuming that the ramp up signals shown in numerical simulations are a principal gravitational wave signal after supernova explosion, one can observationally extract the time evolution of the PNS average density via the direct observation of the gravitational waves by using the fitting formula we found in this study. This is an important information for constraining the EOS for a high density region. We also confirmed that the $f$- and $g_1$-mode frequencies are almost independent of the selection of the PNS surface density in the later phase, i.e., after $\sim 0.3$ sec after core bounce, although the eigenfrequencies of the PNSs generally depend strongly on the selection of the surface density. Furthermore, we pointed out the possibility that the avoided crossing may appear even in the previous numerical simulations. 

In the end, we have to mention a defect in this study. First, in this study, we simply adopt the relativistic Cowling approximation. But, for cold neutron stars, it is known that the eigenfrequencies calculated with the Cowling approximation can deviate from those with full perturbations (without the approximation), e.g., the $f$-mode frequency has as large as $\sim 20\%$ error \citep{YK97}, although one can qualitatively discuss the behavior of the eigenfrequencies. Second, the numerical simulations, whose gravitational wave signals are compared to the frequencies obtained by linear analysis, have been done in Newtonian gravity with the phenomenological general relativistic effect. That is, the gravitational wave signals appearing in the numerical simulations in general relativity may deviate from those considered in this study. Anyway, in order to verify our conclusion in this study, we have to make a linear analysis with full perturbations and compare the calculated eigenfrequencies to the gravitational wave signals obtained via numerical simulation in general relativity. Such a study will be done in the future.

%\newpage
%%%%%%%%%%%%%%%%%%%%%%%%%%%%%%%%%%%%%%%%%%%%%%%%
%\acknowledgments
%%%%%%%%%%%%%%%%%%%%%%%%%%%%%%%%%%%%%%%%%%%%%%%%
\section*{Acknowledgements}
This work is supported in part by Japan Society for the Promotion of Science (JSPS) KAKENHI Grant Numbers  
JP17K05458, % Sotani
JP17H01130, % Kotake-san Kiban A  (Takiwaki Co-PI)
JP17K14306, % Takiwaki-san Wakate B 
JP18H01212, % Yokoi Kinban B (Takiwaki Co-PI)
JP19KK0354, % Sotani 
Ministry of Education, Science and Culture of Japan (MEXT) KAKENHI Grant Numbers 
JP17H06357, % Shingakujutu GWGEN Sokatsu
JP17H06364, % Shingakujutu GWGEN C01,  (Takiwaki Co-PI)
JP17H05206, % Takiwaki-san Shingakujutu-Koubo (Chikasokaku-C02)
JP20H04753. % Sotani Shingakujutu-Koubo
This research is also supported by MEXT as “Program for Promoting Researches on the Supercomputer Fugaku” (Toward a unified view of the universe: from large scale structures to planets), 
Joint Institute for Computational Fundamental Science (JICFuS),
and the National Institutes of Natural Sciences (NINS) program for cross-disciplinary
study (Grant Numbers 01321802 and 01311904) on Turbulence, Transport,
and Heating Dynamics in Laboratory and Solar/Astrophysical Plasmas:
``SoLaBo-X".
Numerical computations were in part carried out on Cray XC50, PC cluster and analysis server at Center for Computational Astrophysics, National Astronomical Observatory of Japan.

\section*{Data availability}
The data underlying this article will be shared on reasonable request to the corresponding author.

\appendix
%%%%%%%%%%%%%%%%%%%%%%%%%%%%%%%%%%%%%%%%%%%%%%%%
\section{Gravitational wave spectra in early phase after core bounce}   % Appendix A
\label{sec:appendix_1}
%%%%%%%%%%%%%%%%%%%%%%%%%%%%%%%%%%%%%%%%%%%%%%%%

\begin{figure*}
\begin{center}
\begin{tabular}{cc}
\includegraphics[scale=0.5]{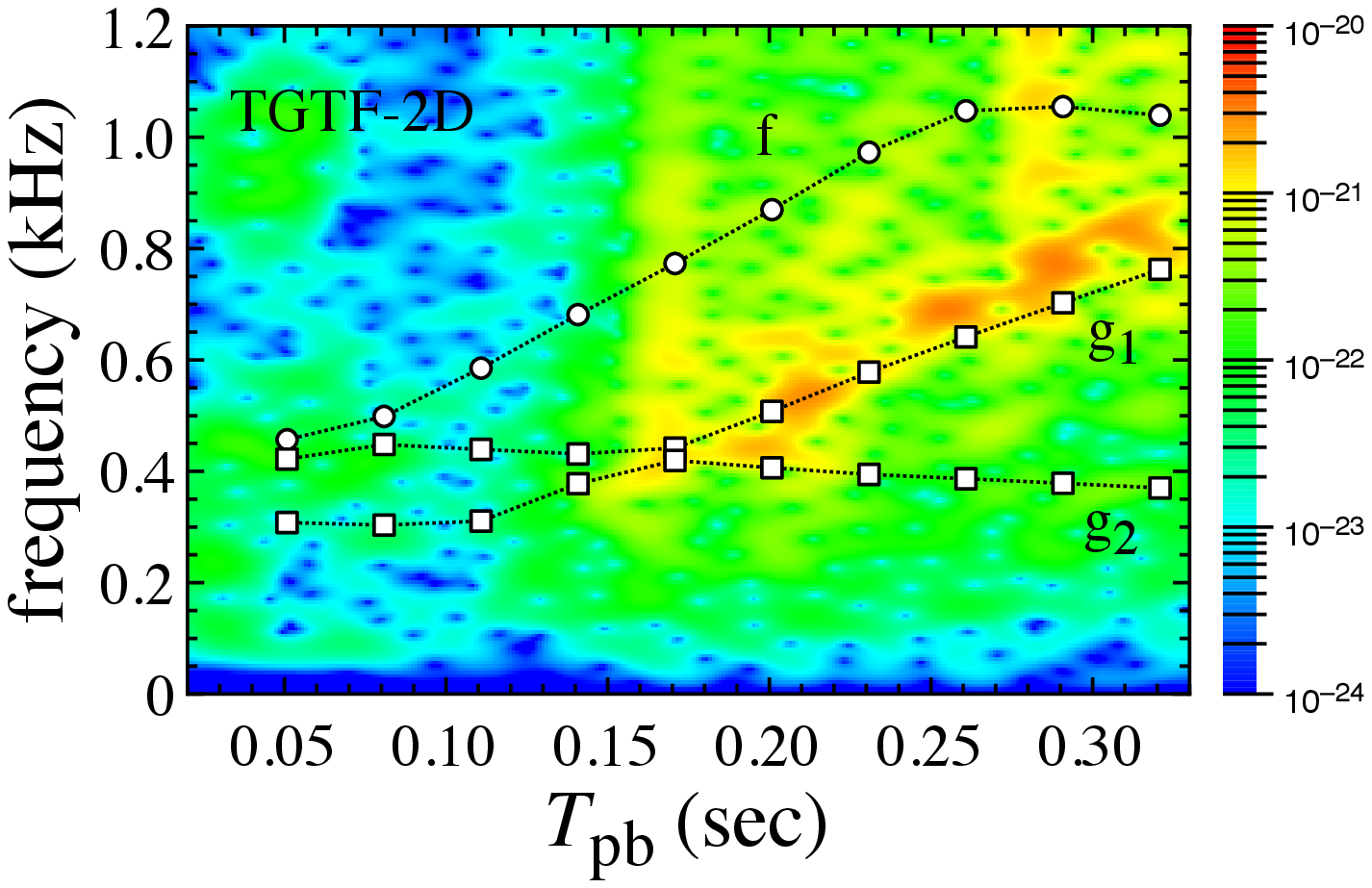} &
\includegraphics[scale=0.5]{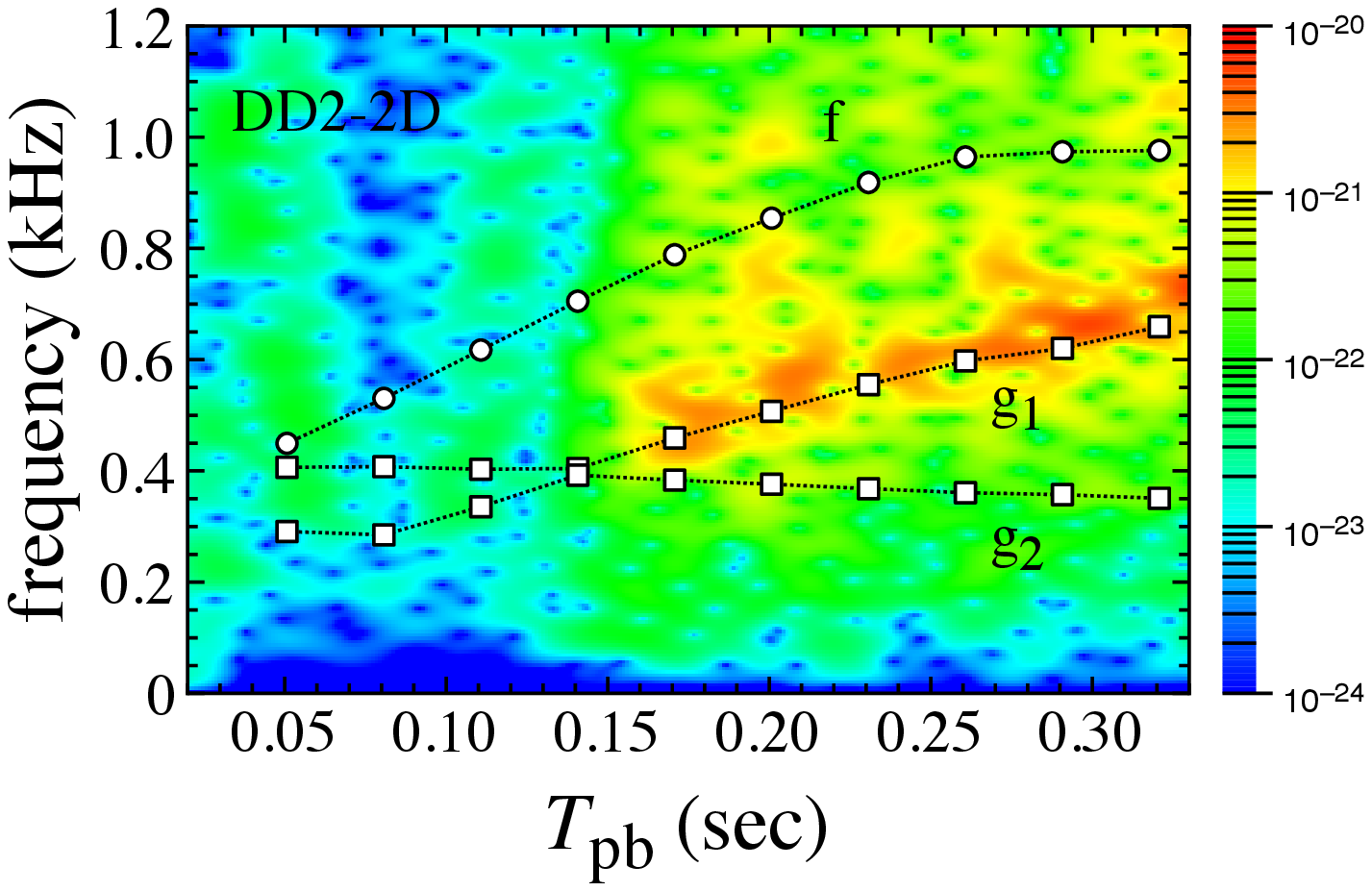}  
\end{tabular}
\end{center}
\caption{%%
Comparison between the gravitational wave signals obtained with the 2D numerical simulation (background contour) and a few eigenfrequencies determined via PNS asteroseismology (marks), where circles and squares denote the $f$- and $g_i$-modes for $i=1$ or 2, respectively, and the left and right panels correspond to the results for TGTF and DD2. The contour denotes the dimensionless gravitational wave strain calculated via Eq. (\ref{eq:strain}), where the source distance is assumed to be $D=10$ kpc.
}%%
\label{fig:2D-spect}
\end{figure*}

\begin{figure}
\begin{center}
\includegraphics[scale=0.5]{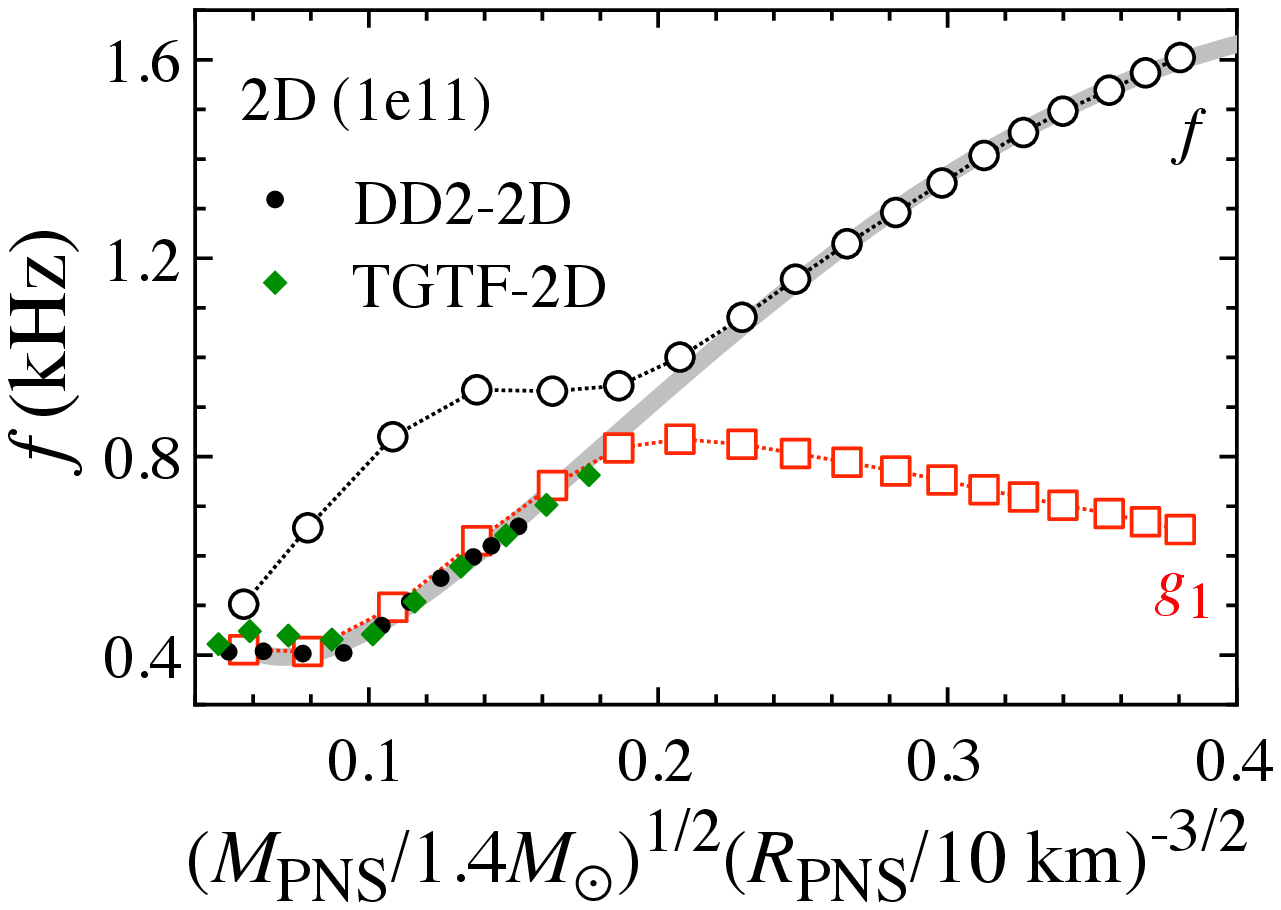}  
\end{center}
\caption{%%
Same as the left panel of Fig. \ref{fig:ft-fit}, but additionally with the data obtained from the PNS models with DD2 (filled circles) and TGTF (filled diamonds), where the thick-solid line is given by Eq. (\ref{eq:fit}).
}%%
\label{fig:ft-fita}
\end{figure}

In this appendix, we show additional results about the comparison between the gravitational wave signals obtained with the 2D numerical simulation and the eigenfrequencies determined via PNS asteroseismology with using the data obtained in the previous our study \citep{ST2020}, even though they are only for early phase after core bounce. In the similar way to the current study, the PNS models are produced with using the data obtained via numerical simulations, adopting two different EOSs with  $20M_\odot$ progenitor model \citep{WH07}. One is the EOS derived with variational method together with Thomas-Fermi approximation for lower density region (referred to as TGTF) \citep{Togashi17}, while another is DD2 based on the relativistic mean field theory \citep{DD2} (see \cite{ST2020} for details). The PNS surface is determined with $\rho_s=10^{11}$ g/cm$^3$. In Fig. \ref{fig:2D-spect}, we show the comparison between the spectra and eigenfrequencies, where the left and right panels correspond to the results for TGTF and DD2, respectively. From this figure, one can clearly see that the gravitational wave signals appearing in the numerical simulation correspond well to the $g_1$-mode frequency for PNSs with $\rho_s=10^{11}$ g/cm$^3$ in early phase after core bounce. In addition, for TGTF model, one may see that the gravitational wave signal comes from the $g_2$-mode frequency, which turns to the $g_1$-mode via the avoided crossing at $\sim 0.17$ sec after core bounce. Furthermore, Fig. \ref{fig:ft-fita} is the figure, where the results for DD2 (filled circles) and for TGTF (filled diamonds) are additionally put in Fig. \ref{fig:ft-fit}. From this figure, one can also clearly see that the frequency of the $g_1$-modes can be expressed well by Eq. (\ref{eq:fit}) independently of the EOS, even thought the progenitor model is completely different from the current study.

%%%%%%%%%%%%%%%%%%%%%%%%%%%%%%%%%%%
% Figure A1
%%%%%%%%%%%%%%%%%%%%%%%%%%%%%%%%%%%
%%%%%%%%%%%%%%%%%%%%%%%%%%%%%%%%%%%

%%%%%%%%%%%%%%%%%%%%%%%%%%%%%%%%%%%
% Figure A2
%%%%%%%%%%%%%%%%%%%%%%%%%%%%%%%%%%%
%%%%%%%%%%%%%%%%%%%%%%%%%%%%%%%%%%%

%%%%%%%%%%%%%%%%%%%%%%%%%%%%%%%%%%%%%%%%%%%%%%%%
\section{Dimensionality dependence of numerical simulations on eigenfrequencies}   % Appendix B
\label{sec:appendix_2}
%%%%%%%%%%%%%%%%%%%%%%%%%%%%%%%%%%%%%%%%%%%%%%%%

\begin{figure}
\begin{center}
\includegraphics[scale=0.5]{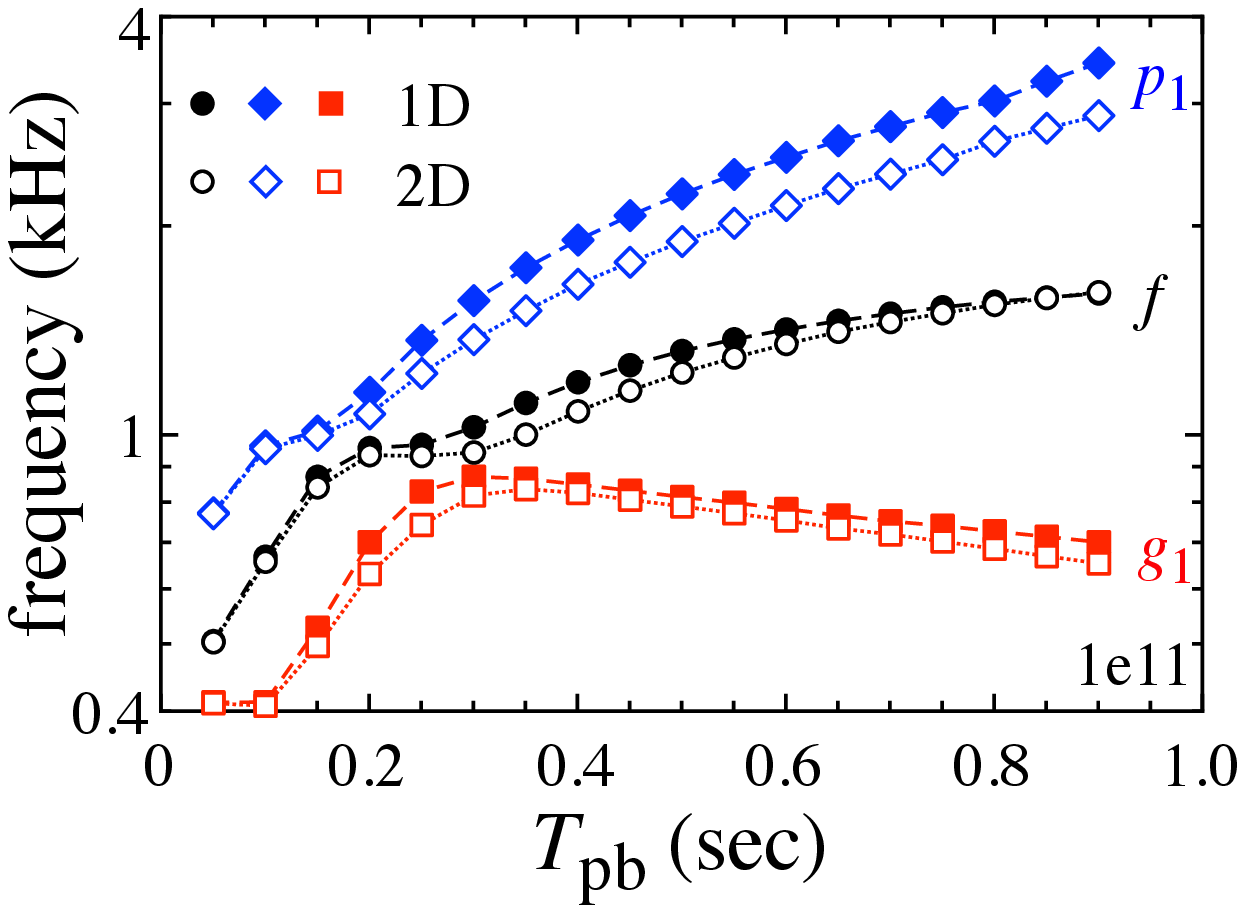}  
\end{center}
\caption{%%
Time evolution of the $f$-, $p_1$-, and $g_1$-modes for the PNSs obtained from the 1D and 2D simulations with $\rho_s=10^{11}$ g/cm$^3$, where the results from 1D and 2D simulations denote with filled and open marks, respectively.
}%%
\label{fig:ft-12D1e11}
\end{figure}

\begin{figure*}
\begin{center}
\begin{tabular}{ccc}
\includegraphics[scale=0.42]{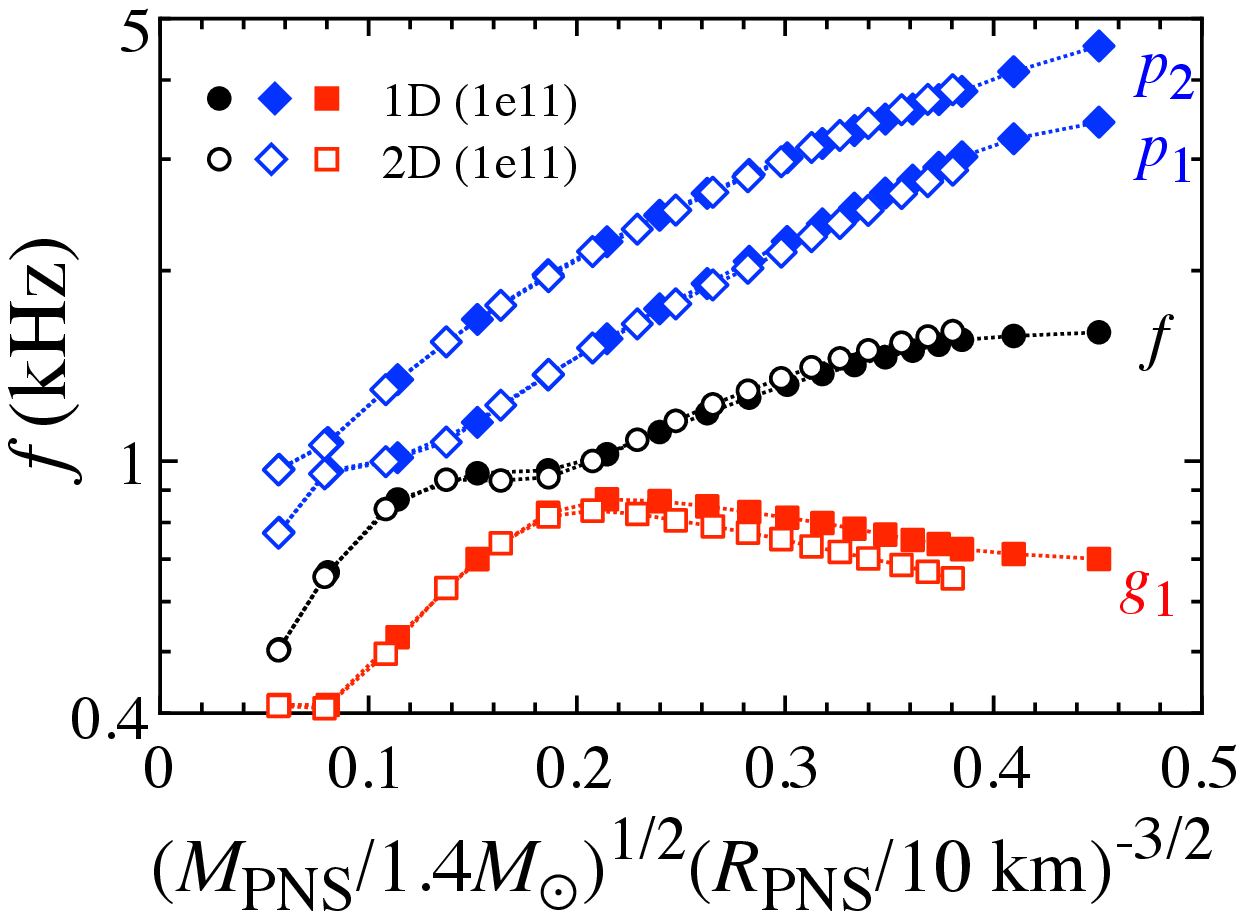} &  
\includegraphics[scale=0.42]{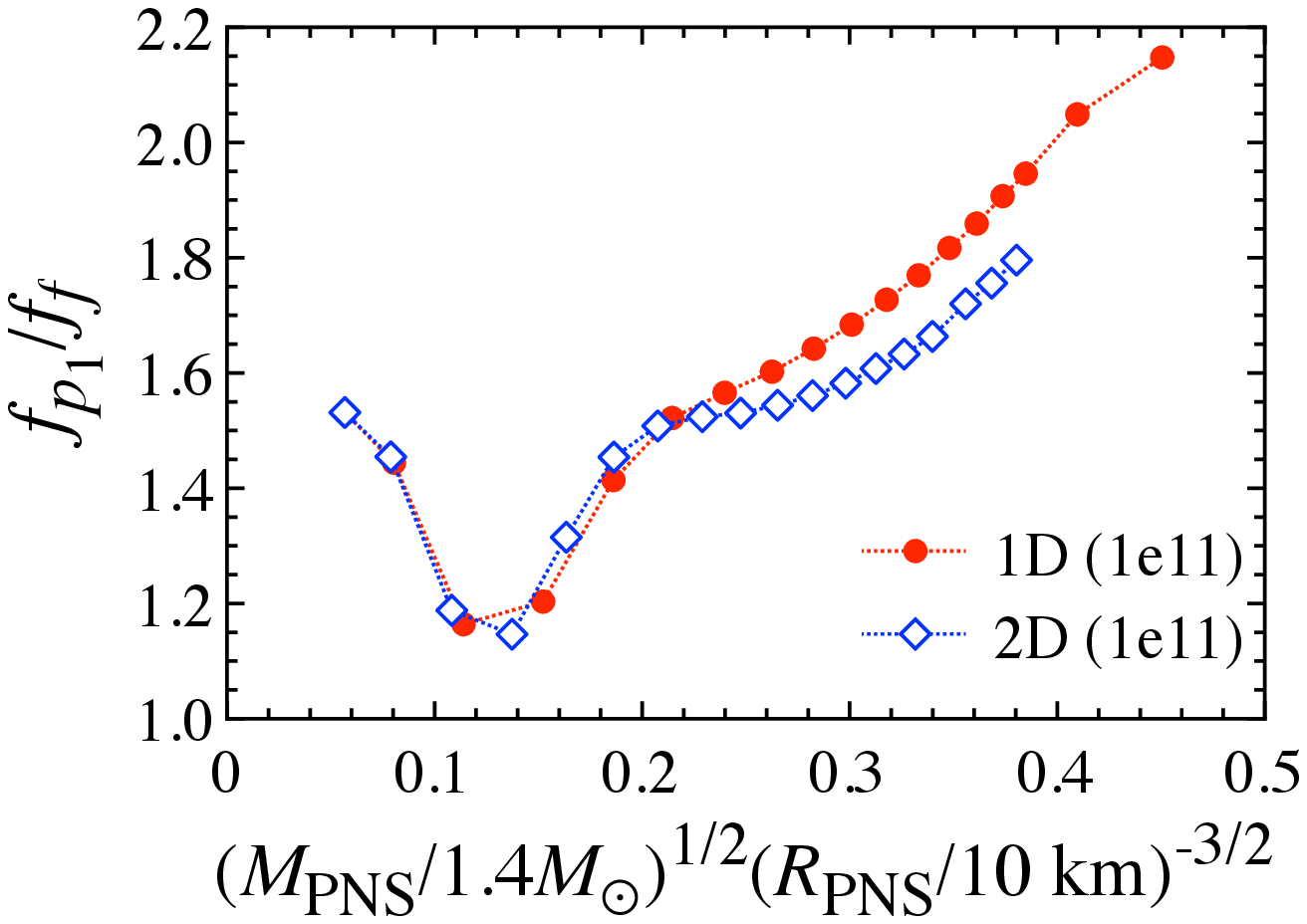} &
\includegraphics[scale=0.42]{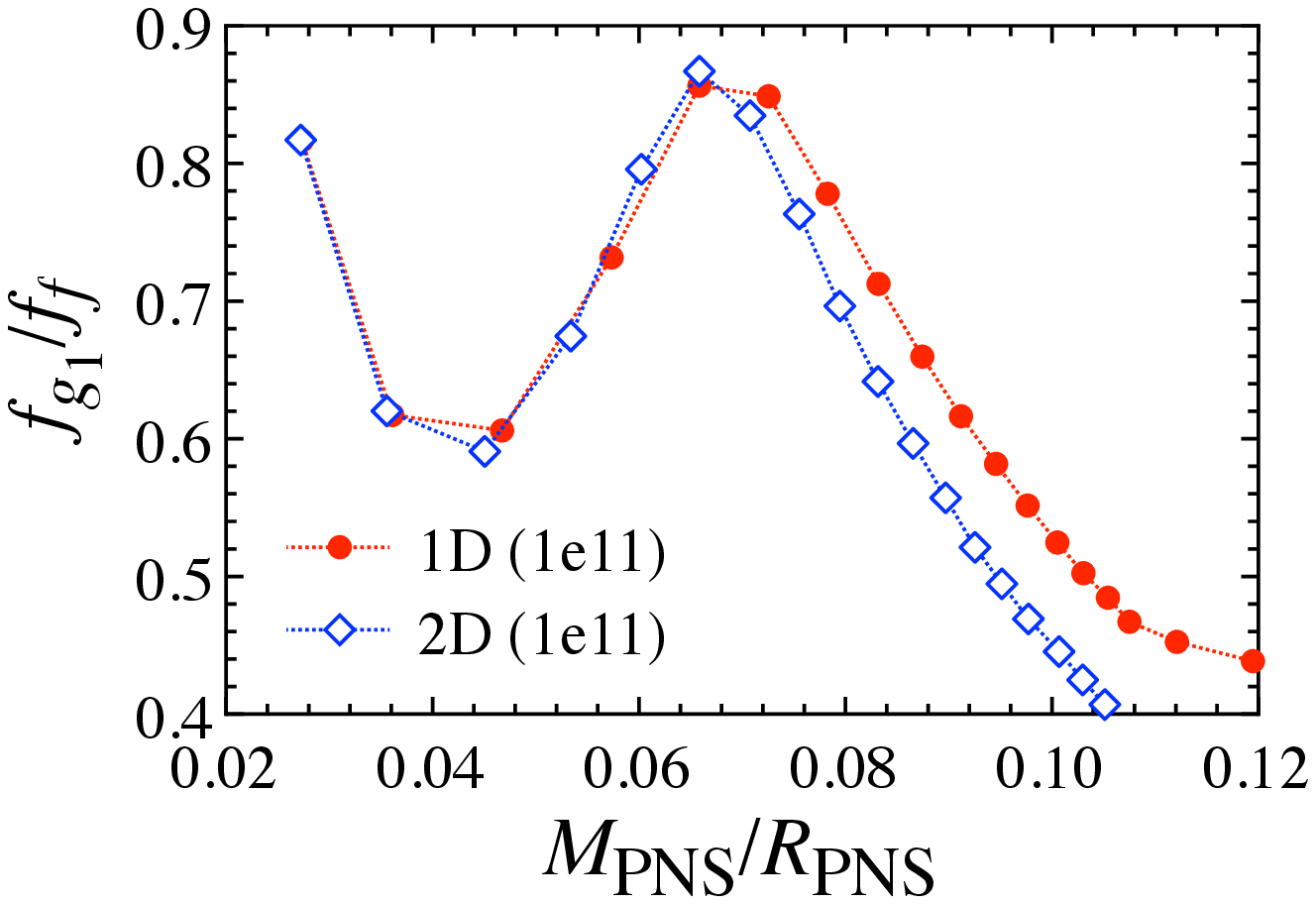} 
\end{tabular}
\end{center}
\caption{%%
Dependence of the eigenmodes on the square root of the PNS average density (left panel), the ratio of the $p_1$-mode to the $f$-mode as a function of the square root of the PNS average density (middle panel), and the ratio of the $g_1$-mode to the $f$-mode as a function of the PNS compactness (right panel), where the filled and open marks respectively denote the results for the PNS models obtained from the 1D and 2D numerical simulations. 
}%%
\label{fig:ff-pf-gf}
\end{figure*}

By using the results discussed in the current study, we also check how the eigenfrequencies depend on the dimensionality of numerical simulations, with which the PNS models are provided, in the similar way to the study in \cite{ST2020}. In Fig. \ref{fig:ft-12D1e11}, we show the time evolution of the $f$-, $p_1$-, and $g_1$-mode frequencies for the PNS models obtained from 1D (filled marks) and 2D simulations (open marks). As pointed out in \cite{ST2020}, the time evolution itself depends on the dimensionality of the numerical simulations. On the other hand, as shown in the left panel of Fig. \ref{fig:ff-pf-gf}, the $f$- and $p_i$-modes weakly depend on the dimensionality as a function of the square root of the stellar average density. Moreover, as pointed out in \cite{ST2020}, one can see from the middle and right panels of Fig. \ref{fig:ff-pf-gf} that at least in the early phase after core bounce the ratio of the $p_1$-mode to the $f$-mode is almost independent of the dimensionality as a function of the square root of the average density, while the ratio of the $g_1$-mode to the $f$-mode is almost independent of the dimensionality as a function of the compactness. Even so, one can also see a deviation in such a ratio of eigenmodes after the avoided crossing between the $f$- and the $g_1$-modes, where the ratio of the $g_1$-mode to the $f$-mode becomes maximum. This deviation may come from the qualitative difference of the PNS mass evolution, depending on the dimensionality of numerical simulation as shown in Fig. \ref{fig:MRt}.

%%%%%%%%%%%%%%%%%%%%%%%%%%%%%%%%%%%
% Figure B1
%%%%%%%%%%%%%%%%%%%%%%%%%%%%%%%%%%%
%%%%%%%%%%%%%%%%%%%%%%%%%%%%%%%%%%%

%%%%%%%%%%%%%%%%%%%%%%%%%%%%%%%%%%%
% Figure B2
%%%%%%%%%%%%%%%%%%%%%%%%%%%%%%%%%%%
%%%%%%%%%%%%%%%%%%%%%%%%%%%%%%%%%%%

%%%%%%%%%%%%%%%%%%%%%%%%%%%%%%%%%%%%%%%%%%%%%%%%


\begin{thebibliography}{999}
%%%%%%%%%%%%%%%%%%%%%%%%%%%%%%%%%%%%%%%%%%%%%%%%

\bibitem[\protect\citeauthoryear{Abbott et al.}{2017a}]{GW6}  % GW170817
   Abbott B. P. et al. (LIGO Scientific Collaboration and Virgo Collaboration), 2017, Phys. Rev. Lett., 119, 161101

\bibitem[\protect\citeauthoryear{Abbott et al.}{2017b}]{EM}  % EM in GW170817
   Abbott B. P. et al. (LIGO Scientific Collaboration and Virgo Collaboration), 2017, ApJ, 848, L12

\bibitem[\protect\citeauthoryear{Abbott et al.}{2017c}]{CE}
   Abbott B. P. et al. (LIGO Scientific Collaboration and Virgo Collaboration), 2017, Class. Quantum Grav., 34, 044001

\bibitem[\protect\citeauthoryear{Andresen et al.}{2017}]{Andresen16} 
    Andresen H., M\"{u}ller B., M\"{u}ller E., Janka H.-T., 2017, MNRAS, 468, 2032

\bibitem[\protect\citeauthoryear{Andersson \& Kokkotas}{1996}]{AK1996}
   Andersson N., Kokkotas K. D., 1996, Phys.\ Rev.\ Lett., 77, 4134

\bibitem[\protect\citeauthoryear{Andersson \& Kokkotas}{1998}]{AK1998}
   Andersson N., Kokkotas K. D., 1998, MNRAS, 299, 1059

\bibitem[\protect\citeauthoryear{Aso et al.}{2013}]{aso13} 
   Aso Y., Michimura Y., Somiya K., Ando M., Miyakawa O., Sekiguchi T., Tatsumi D., Yamamoto H., 2013, Phys. Rev. D, 88, 043007

\bibitem[\protect\citeauthoryear{Burgio et al.}{2011}]{Burgio2011}
   Burgio G. F., Ferrari V., Gualtieri L., Schulze H.-J., 2011, Phys.\ Rev.\ D, 84, 044017

\bibitem[\protect\citeauthoryear{Camelio et al.}{2017}]{Camelio17}
   Camelio G., Lovato A., Gualtieri L., Benhar O., Pons J. A., Ferrari V., 2017, Phys. Rev. D, 96, 043015

%\bibitem[\protect\citeauthoryear{Cerd\'{a}-Dur\'{a}n., Stergioulas, \& Font}{2009}]{CSF2009}
%   Cerd\'{a}-Dur\'{a}n P., Stergioulas N., Font J. A., 2009 MNRAS, 397, 1607

\bibitem[\protect\citeauthoryear{Cerd\'{a}-Dur\'{a}n et al.}{2013}]{CDAF2013}
  Cerd\'{a}-Dur\'{a}n P., DeBrye N., Aloy M. A., Font J. A., Obergaulinger M., 2013, ApJ, 779, L18

\bibitem[\protect\citeauthoryear{Colaiuda \& Kokkotas}{2012}]{CK2012}
   Colaiuda A., Kokkotas K. D., 2012, MNRAS, 423, 811

\bibitem[\protect\citeauthoryear{Doneva et al.}{2013}]{DGKK2013}
   Doneva D. D., Gaertig E., Kokkotas K. D., Kr\"{u}ger C., 2013, Phys.\ Rev.\ D, 88, 044052

\bibitem[\protect\citeauthoryear{Ferrari, Miniutti, \& Pons}{2003}]{FMP2003}
   Ferrari V., Miniutti G., Pons J. A., 2003, MNRAS, 342, 629

\bibitem[\protect\citeauthoryear{Fuller et al.}{2015}]{FKAO2015}
   Fuller J., Klion H., Abdikamalov E., Ott C. D., 2015, MNRAS, 450, 414

\bibitem[\protect\citeauthoryear{Gabler et al.}{2011}]{GCFMS2011}
   Gabler M., Cerd\'{a}-Dur\'{a}n P., Font J. A., M\"{u}ller E., Stergioulas N., 2011, MNRAS, 410, L37

\bibitem[\protect\citeauthoryear{Gabler et al.}{2013}]{GCFMS2013b}
   Gabler M., Cerd\'{a}-Dur\'{a}n P., Stergioulas N., Font J. A., Muller E., 2013, Phys.\ Rev.\ Lett., 111, 211102
%  Imprints of Superfluidity on Magnetoelastic Quasiperiodic Oscillations of Soft Gamma-Ray Repeaters

\bibitem[\protect\citeauthoryear{Gearheart et al.}{2011}]{GNHL2011}
  Gearheart M., Newton W. G., Hooker J., Li B. A., 2011, MNRAS, 418, 2343

\bibitem[\protect\citeauthoryear{Kawahara et al.}{2018}]{Kawahara18}
  Kawahara H., Kuroda T., Takiwaki T., Hayama K., Kotake K., 2018, ApJ, 867, 126
 
\bibitem[\protect\citeauthoryear{Kokkotas \& Schmidt}{1999}]{KS1999}
   Kokkotas K. D., Schmidt B. G., 1999, Living Rev. Relativ., 2, 2

\bibitem[\protect\citeauthoryear{Kotake et al.}{2015}]{kotake2018}
   Kotake K., Takiwaki T., Fischer T., Nakamura K., Mart\'{i}nez-Pinedo G., 2018, ApJ, 853, 170

\bibitem[\protect\citeauthoryear{Kuroda, Kotake, \& Takiwaki}{2016}]{KKT2016}
   Kuroda T., Kotake K., Takiwaki T., 2016, ApJ, 829, L14

\bibitem[\protect\citeauthoryear{Lattimer \& Swesty}{1991}]{LS220}
  Lattimer J. M., Swesty F. D., 1991, Nucl. Phys. A 535, 331

\bibitem[\protect\citeauthoryear{Liebendoerfer, Whitehouse, \& Fischer}{2009}]{liebendoerfer2009}
  Liebendoerfer M., Whitehouse S., Fischer T., 2009, ApJ, 698, 1174

\bibitem[\protect\citeauthoryear{Moriya et al.}{2019}]{He29} % He2.9
  Moriya T. J., Mazzali P. A., Tominaga N., Hachinger S., Blinnikov S. I., Tauris T. M., Takahashi K., Tanaka M., Langer N., Podsiadlowski P., 2017, MNRAS, 466, 2085

\bibitem[\protect\citeauthoryear{Morozova et al.}{2018}]{MRBV2018}
  Morozova V., Radice D., Burrows A., Vartanyan D., 2018, ApJ, 861, 10

\bibitem[\protect\citeauthoryear{M\"{u}ller, Janka, \& Marek}{2013}]{MJM2013}
   M\"{u}ller B., Janka H. -T., Marek A., 2013, ApJ, 766, 43

\bibitem[\protect\citeauthoryear{Murphy, Ott, \& Burrows}{2009}]{Murphy09}
   Murphy J.~W., Ott C.~D., Burrows A., 2009, ApJ, 707, 1173

\bibitem[\protect\citeauthoryear{Nakamura, Takiwaki, \& Kotake}{2019}]{nakamura2019}
   Nakamura K., Takiwaki T., Kotake K., 2019, PASJ, 71, 98

\bibitem[\protect\citeauthoryear{O'Connor \& Couch}{2018}]{OC2018}
   O'Connor E. P., Couch S. M., 2018, ApJ, 865, 81
   
\bibitem[\protect\citeauthoryear{O'Connor et al.}{2018}]{oconnor2018}
  O'Connor E. {\it et~al.}, 2018, J. Phys. G, 45, 104001

\bibitem[\protect\citeauthoryear{Ott et al.}{2013}]{Ott13} 
   Ott, C.~D. Abdikamalov E., M{\"o}sta P., Haas R., Drasco S., O'Connor E. P., Reisswig C., Meakin C. A., Schnetter E., 2013, ApJ 768, 115

\bibitem[\protect\citeauthoryear{Passamonti \& Andersson}{2012}]{PA2012}
   Passamonti A., Andersson N., 2012, MNRAS, 419, 638
% Towards real neutron star seismology: Accounting for elasticity and superfluidity

\bibitem[\protect\citeauthoryear{Powell \& M\"{u}ller}{2020}]{PM20}
  Powell J., M\"{u}ller B., arXiv:2002.10115

\bibitem[\protect\citeauthoryear{Punturo, L\"{u}ck, \& Beker}{2014}]{punturo} 
   Punturo M., L\"{u}ck H., Beker M., {\it A Third Generation Gravitational Wave Observatory: The Einstein Telescope},
   edited by M. Bassan, Advanced Interferometers and the Search for Gravitational Waves. Astrophysics and Space
   Science Library Vol. 404 (Springer, Cham, 2014)

\bibitem[\protect\citeauthoryear{Radice et al.}{2019}]{RMBVN19}
   Radice D., Morozova V., Burrows A., Vartanyan D., Nagakura H., 2019, ApJ, 876, L9

\bibitem[\protect\citeauthoryear{Richers et al.}{2017}]{Richers2017}
   Richers S., Ott C.~D., Abdikamalov E., O'Connor E., Sullivan C., 2017, Phys. Rev. D, 95, 063019

\bibitem[\protect\citeauthoryear{Sasaki et al.}{2019}]{sasaki2019}
   Sasaki H., Takiwaki T., Kawagoe S., Horiuchi S., Ishidoshiro K., arXiv:1907.01002.

\bibitem[\protect\citeauthoryear{Sotani, Tominaga \& Maeda}{2001}]{STM2001}
   Sotani H., Tominaga K., Maeda K. I., 2001, Phys.\ Rev.\ D, 65, 024010
  
\bibitem[\protect\citeauthoryear{Sotani, Kohri \& Harada}{2004}]{SKH2004}
   Sotani H., Kohri K., Harada T., 2004, Phys.\ Rev.\ D, 69, 084008

\bibitem[\protect\citeauthoryear{Sotani, Kokkotas \& Stergioulas}{2007}]{SKS2007}
   Sotani H., Kokkotas K. D., Stergioulas N., 2007, MNRAS, 375, 261

\bibitem[\protect\citeauthoryear{Sotani, Kokkotas \& Stergioulas}{2008}]{SKS2008}
   Sotani H., Kokkotas K. D., Stergioulas N., 2008a, MNRAS, 385, L5

\bibitem[\protect\citeauthoryear{Sotani et al.}{2011}]{SYMT2011}
   Sotani H., Yasutake N., Maruyama T., Tatsumi T., 2011, Phys.\ Rev.\ D, 83, 024014

\bibitem[\protect\citeauthoryear{Sotani et al.}{2012}]{SNIO2012}
   Sotani H., Nakazato K., Iida K., Oyamatsu K., 2012, Phys.\ Rev.\ Lett., 108, 201101

\bibitem[\protect\citeauthoryear{Sotani et al.}{2013a}]{SNIO2013a}   
   Sotani H., Nakazato K., Iida K., Oyamatsu K., 2013a, MNRAS, 428, L21

\bibitem[\protect\citeauthoryear{Sotani et al.}{2013b}]{SNIO2013b}   
   Sotani H., Nakazato K., Iida K., Oyamatsu K., 2013b, MNRAS, 434, 2060

\bibitem[\protect\citeauthoryear{Sotani, Iida \& Oyamatsu}{2016}]{SIO2016} 
   Sotani H., Iida K., Oyamatsu K., 2016, New Astron., 43, 80

\bibitem[\protect\citeauthoryear{Sotani \& Takiwaki}{2016}]{ST2016}
   Sotani H., Takiwaki T., 2016, Phys.\ Rev.\ D, 94, 044043

\bibitem[\protect\citeauthoryear{Sotani, Iida \& Oyamatsu}{2017}]{SIO2017} 
   Sotani H., Iida K., Oyamatsu K., 2017, MNRAS, 464, 3101

\bibitem[\protect\citeauthoryear{Sotani et al.}{2017}]{SKTK2017}
   Sotani H., Kuroda T., Takiwaki T., Kotake K., 2017, Phys.\ Rev.\ D, 96, 063005

\bibitem[\protect\citeauthoryear{Sotani, Iida \& Oyamatsu}{2018}]{SIO2018} 
   Sotani H., Iida K., Oyamatsu K., 2018, MNRAS, 479, 4735

\bibitem[\protect\citeauthoryear{Sotani, Iida \& Oyamatsu}{2019}]{SIO2019} 
   Sotani H., Iida K., Oyamatsu K., 2019, MNRAS, 489, 3022

\bibitem[\protect\citeauthoryear{Sotani et al.}{2019}]{SKTK2019}
   Sotani H., Kuroda T., Takiwaki T., Kotake K., 2019, Phys. Rev. D, 99, 123024

\bibitem[\protect\citeauthoryear{Sotani \& Sumiyoshi}{2019}]{SS2019}
   Sotani H., Sumiyoshi K., 2019, Phys.\ Rev.\ D, 100, 083008

\bibitem[\protect\citeauthoryear{Sotani \& Takiwaki }{2020}]{ST2020}
   Sotani H., Takiwaki T., 2020, Phys.\ Rev.\ D, 102, 023028

\bibitem[\protect\citeauthoryear{Steiner, Hempel, \& Fischer}{2013}]{SHF2013} % SFHx EOS
   Steiner A. W., Hempel M., Fischer T., 2013, ApJ, 774, 17

\bibitem[\protect\citeauthoryear{Takiwaki, Kotake, \& Suwa}{2014}]{takiwaki2014}
  Takiwaki T., Kotake K., Suwa Y., 2014, ApJ, 786, 83

\bibitem[\protect\citeauthoryear{Takiwaki, Kotake, \& Suwa}{2016}]{takiwaki2016}
  Takiwaki T., Kotake K., Suwa Y., 2016, MNRAS, 461, L112

\bibitem[\protect\citeauthoryear{Takiwaki \& Kotake}{2018}]{Takiwaki2017}
    Takiwaki T., Kotake K., 2018, MNRAS, 475, L91

\bibitem[\protect\citeauthoryear{Takiwaki}{2020a}]{takiwaki2020usm}
  Takiwaki T., in preparation

\bibitem[\protect\citeauthoryear{Takiwaki}{2020b}]{takiwaki2020eos}
  Takiwaki T., in preparation

\bibitem[\protect\citeauthoryear{Togashi et al.}{2017}]{Togashi17}  % TGTF
   Togashi H., Nakazato K., Takehara Y., Yamamuro S., Suzuki H., Takano M., 2017, Nucl. Phys. A 961, 78

\bibitem[\protect\citeauthoryear{Torres-Forn\'{e} et al.}{2018}]{TCPF2018}
   Torres-Forn\'{e} A., Cerd\'{a}-Dur\'{a}n P., Passamonti A., Font J. A., 2018, MNRAS, 474, 5272

\bibitem[\protect\citeauthoryear{Torres-Forn\'{e} et al.}{2019a}]{TCPOF2019a}
   Torres-Forn\'{e} A., Cerd\'{a}-Dur\'{a}n P., Passamonti A., Obergaulinger M., Font J. A., 2019a, MNRAS, 482, 3967

\bibitem[\protect\citeauthoryear{Torres-Forn\'{e} et al.}{2019b}]{TCPOF2019b}
   Torres-Forn\'{e} A., Cerd\'{a}-Dur\'{a}n P., Obergaulinger M., M\"{u}ller B., Font J. A., 2019, Phys. Rev. Letts., 123, 051102

\bibitem[\protect\citeauthoryear{Typel et al.}{2010}]{DD2}  % DD2
   Typel S., Ropke G., Klahn T., Blaschke D., Wolter H. H., 2010, Phys. Rev. C, 81, 015803
%   T. Fischer, M. Hempel, I. Sagert, Y. Suwa, and J. Schaffner-Bielich, Eur. Phys. J. A {\bf 50}, 46 (2014).

\bibitem[\protect\citeauthoryear{Vartanyan, Burrows, \& Radice}{2019}]{VBR2019}
   Vartanyan D., Burrows A., Radice D., 2019, MNRAS, 489, 2227

\bibitem[\protect\citeauthoryear{Westernacher-Schneider et al.}{2019}]{WS2019}
   Westernacher-Schneider J. R., O'Connor E., O'Sullivan E., Tamborra I., Wu M.-R., Couch S. M., Malmenbeck F., 2019,  
   Phys. Rev. D, 100, 123009

\bibitem[\protect\citeauthoryear{Woosley \& Weaver}{1995}]{WW95}  % progenitor 15M_sun for \cite{KKT2016}
   Woosley S. E., Weaver T. A., 1995,  ApJ. Suppl., 101, 181

\bibitem[\protect\citeauthoryear{Woosley \& Heger}{2007}]{WH07}  % WH07
   Woosley S. E., Heger A., 2007, Phys. Rep., 442, 269

\bibitem[\protect\citeauthoryear{Yakunin et al.}{2015}]{Yakunin15}
    Yakunin K.~N., Mezzacappa A., Marronetti P., Yoshida S., Bruenn S.~W., 
    Hix W.~R., Lentz E.~J., Bronson Messer O.~E., Harris J.~A., Endeve E.,
    Blondin J.~M., Lingerfelt E.~J., 2015, Phys.\ Rev.\ D, 92, 084040

\bibitem[\protect\citeauthoryear{Yoshida \& Kojima}{1997}]{YK97}   % accuracy of Cowling
   Yoshida S., Kojima Y., 1997, MNRAS 289, 117

\bibitem[\protect\citeauthoryear{Zaizen et al.}{2019}]{zaizen2019}
   Zaizen M., Cherry J. F., Takiwaki T., Horiuchi S., Kotake K., Umeda H., Yoshida T., arXiv:1908.10594.



\end{thebibliography}
\end{document}